\documentclass{aa}  
\usepackage{amsmath}
\usepackage{graphicx}
\usepackage{txfonts}
\usepackage{natbib} %[authoryear]{natbib}
\usepackage{subfig}

\begin{document}
\def\mean#1{\left< #1 \right>}

   \title{Establishing binarity amongst Galactic RV Tauri stars with a disc
   \thanks{Based on observations made with the Flemish Mercator Telescope and the Swiss Leonhard Euler Telescope.}} %\fnmsep}
   \author{Rajeev Manick\inst{1}
          \and
          Hans Van Winckel\inst{1}
          \and
          Devika Kamath\inst{1}
          \and
          Michel Hillen\inst{1}
          \and
          Ana Escorza \inst{1,2}
          }

      \institute{Instituut voor Sterrenkunde (IvS), KU Leuven,
              Celestijnenlaan 200D, B-3001 Leuven, Belgium\\
              \email{rajeev@ster.kuleuven.be}
            \and
            Institut d'Astronomie et d'Astrophysique, Universit\'e Libre de
Bruxelles, CP 226, Boulevard du Triomphe, B-1050 Bruxelles, Belgium
             }

   \authorrunning{Manick et al. 2016}
   \titlerunning{Binarity amongst RV Tauri stars with a disc}

% \abstract{}{}{}{}{} 
% 5 {} token are mandatory
   \abstract
   % context heading (optional)
   {During the last few decades, it became more evident that binarity is a prevalent phenomenon amongst RV Tauri stars with a disc. This study is a contribution in comprehending the role of binarity upon late stages of stellar evolution.}
   {In this paper we determine the binary status of six Galactic RV Tauri stars, namely DY\,Ori, EP\,Lyr, HP\,Lyr, IRAS\,17038$-$4815, IRAS\,09144$-$4933 and TW\,Cam, which are surrounded by a dusty disc. The radial velocities are contaminated by high amplitude pulsations. We disentangle the pulsations from the orbital signal in order to determine accurate orbital parameters. We also place them 
   on the HR diagram, thereby establishing their evolutionary nature.}{We used high resolution spectroscopic time series obtained from the HERMES and CORALIE spectrographs mounted on the Flemish
    Mercator and Swiss Leonhard Euler Telescopes, respectively. An updated ASAS/AAVSO photometric time series is analysed to complement 
    the spectroscopic pulsation search and to clean the radial velocities from the pulsations. 
    The pulsation-cleaned orbits are fitted with a Keplerian model to determine the spectroscopic orbital parameters. We also calibrated a PLC relationship using type II cepheids in the LMC 
    and apply the relation to our Galactic sample to obtain accurate distances and hence
    luminosities.}{All the six Galactic RV Tauri stars included in this study are binaries with orbital periods ranging between $\sim$\,650 and 1700 days
    and with eccentricities between 0.2 and 0.6. The mass functions range between 0.08 to 0.55\,M$_{\odot}$ which points to an unevolved low mass companion. In the
    photometric time series we detect a long-term variation on the time-scale of the orbital period for IRAS\,17038$-$4815, IRAS\,09144$-$4933 and TW\,Cam. Our derived stellar luminosities
    indicate that all except DY\,Ori and EP\,Lyr, are post-AGB stars. DY\,Ori and EP\,Lyr are likely examples of the recently discovered dusty post-RGB stars.
    }{The orbital parameters strongly suggest that the evolution of these stars was interrupted by a strong phase of binary interaction during or even prior to the AGB. 
    The observed eccentricities and long orbital periods among these stars provides a challenge to the standard theory of binary evolution.}


   \maketitle
%
%________________________________________________________________curve

\section{Introduction}

RV Tauri stars are a group of luminous radial pulsators that occupy the high luminosity end of the population II cepheid
instability strip. They were classified as post-asymptotic giant branch (post-AGB) stars by \citet{jura86} 
because of their observational properties. They occupy the F, G or K spectral classes, they are highly 
luminous and many of them have an infrared excess (IR-excess) which indicates the presence of circumstellar material.

RV Tauri stars are characterised and defined by their light curves that show alternating deep and shallow minima.
The period between two successive minima is typically between 20 and 75 days and is called the fundamental period. 
The time between two successive deep minima has a range typically twice that of the fundamental 
period and is very often termed the formal period. These stars occupy the higher period range 
among the population II cepheids \citep{wallerstein02}.  As yet, there exists no definite 
characterisation of the photometric wavering related to the pulsations of RV Tauri stars.
Some objects display a very distinct alternation of shallow and deep minima, while for others the lightcurves 
are less stable and are more semi-periodic.
Nevertheless, it is now widely accepted that most of the RV Tauri stars can be sub-divided photometrically in 
two groups based on their long term variability: the RVa and RVb types. The RVa type stars have a constant mean magnitude 
while the RVb types vary slowly in mean magnitude with a long secondary period of around 600 to 2600 days \citep{pollard96}. 

In contrast to the photometric classification based 
primarily on the light variability, a \textit{spectroscopic} classification of RV Tauri stars 
was presented in \citet{preston63}. Three distinct spectroscopic groups were defined: (RVA) those with 
spectral types G or K, (RVB) those with spectral types Fp(R), weak lines in their spectra and strong CN and CH bands, and (RVC) which are like RVB but with little or no trace of CN bands.

\begin{figure}[h]
   \centering
   \includegraphics[width=8cm,height=6cm]{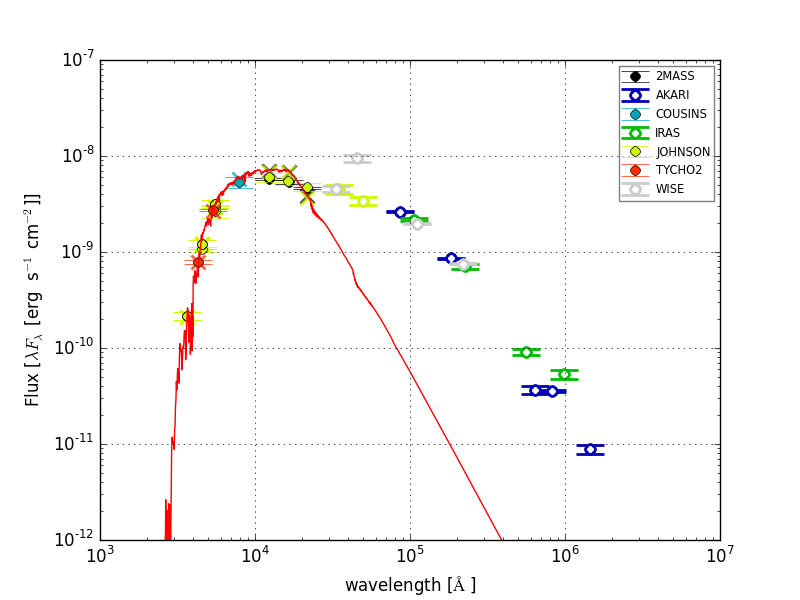}
   \caption{An example of the SED of TW\,Cam; a RV Tauri star surrounded by a disc.
   The red line shows the fit with a photospheric model atmosphere. The clear broad IR-excess can already be seen at a wavelength of $\sim$ 2\,$\mu m$, indicative of a dusty disc.}
     \label{figure:TW_CamSED}
\end{figure}

About 23\% of the 126 RV Tauri stars known in the Galaxy \citep[see, GCVS:][]{samus09},
exhibit an IR-excess in their spectral energy distribution (SED), as illustrated with an example in Fig.~\ref{figure:TW_CamSED}. 
Colour-colour plots made on the basis of infrared surveys like IRAS (Infra-Red Astronomical Satellite) and the WISE (Wide-field Infrared Survey Explorer) have enabled a classification of these excesses.

\citet{lloydevans85}, showed that RV Tauri stars occupy a specific region in 
the IRAS colour-colour plot: \textit{the RV Tauri box}. A more in-depth study by \citet{gezer2015} explores the WISE colour-colour 
plot and compares Galactic RV Tauri stars with a reference sample of Galactic post-AGB stars
in relation to their IR properties, binary status and chemical abundances. They found that a large fraction of the Galactic RV Tauri stars have the same infrared properties as the reference stars with a known disc. \citet{bujarrabal15} and \citet{hillen2015} confirmed that RV Tauri stars which have a broad IR excess, mimicking that of post-AGB stars, indeed have a circumbinary disc by proving the disc-like structure around the RV Tauri star, AC Her. The discs are all found around binary systems; hence these RV Tauri stars are likely binaries as well.

In the past few decades, it became clear that the photospheres
of many post-AGB and RV Tauri stars, show chemical anomalies, the so-called depletion phenomenon. Refractory elements with high condensation 
temperature (e.g, Si, Ti and Fe) are under-abundant, while volatile elements with lower condensation temperature 
(e.g, C, N, O and Zn) are found over-abundant or close to solar abundances \citep{giridhar94,giridhar98,giridhar00,vanwinckel98, maas05}.

A well accepted explanation for this anomaly is given by \citet{waters92}, in terms of selective re-accretion of gas from the circumbinary environment, while the dust remains stable at the sublimation radius as a consequence of radiation pressure. This idea is confirmed by the observation that a large fraction of depleted stars are surrounded by a disc \citep{gezer2015}. \citet{vanwinckel95} showed that the extremely depleted iron-deficient post-AGB stars HR 4049 \citep{lambert88,waelkens91a}, HD 44179, HD 52961 \citep{waelkens91b}, and BD$+$39$^{\circ}$4926 \citep{kodaira70} are actually indeed binaries.

There is growing evidence that binarity is a needed condition for disc formation or 
conversely, a Keplerian disc around a post-AGB system implies that it is a binary. \citet{vanwinckel09} studied 
a sample of six post-AGB stars which are low amplitude pulsators with a disc and found them all to be binaries. 
These provide good indication that the disc-binarity correlation is indeed bona fide among 
post-AGB stars. Is this also the case for RV Tauri stars with disc-like IR excess?

At present, there are 6 confirmed binaries in a sample of 126 known Galactic RV Tauri stars, namely, U Mon \citep{pollard95,pollard97}, IW Car \citep{pollard97},
AC Her \citep{vanwinckel98}, ENTra \citep{vanwinckel99}, SX Cen \citep{maas02} and RU Cen \citep{maas02}. 
The low fraction of confirmed spectroscopic binaries is very likely an observational bias, which is strengthened by the difficulty to trace orbital motion in large-amplitude pulsators.
Another reason is the presence of shocks in the photospheres, causing line-splitting and line profile variability 
in the spectra which again makes the orbit hard to detect.

This paper is a contribution in comprehending the disc-binarity liaison amidst RV Tauri stars. In Sect.~\ref{section:sampleData} we describe the sample of RV Tauri stars presented in this study, together with the photometric and spectroscopic data.
Sect.~\ref{section:photometrypulsations} exhibits our photometric analysis including the pulsations, SEDs and PLC relationship.
The method used for obtaining pulsation-cleaned orbits, leading to the spectroscopic results are described in Sect.~\ref{section:spectroscopy}. In Sect. \ref{section:discussion} we discuss our results and their implications and we conclude in Sect. \ref{section:conclusion}.

%__________________________________________________________________
\section{Sample selection and data overview} \label{section:sampleData}

\subsection{Sample} \label{section:sample}
Of the 126 Galactic RV Tauri stars \citep{samus09}, 23 objects with a near IR-excess, are part of the ongoing HERMES (High Efficiency and 
Resolution Mercator Echelle Spectrograph) radial velocity monitoring program. 
We selected RV Tauri stars from that sample, which are not yet known to be binaries and which had sufficient data coverage for a radial velocity study. 
The mentioned selection criteria resulted in a sample of 6 Galactic RV Tauri 
stars listed in Table \ref{table:sampleselection}.

Since the monitoring programme is carried out by relatively small sized telescopes, 
we are observationally limited towards bright objects with a mean V-mag $\lesssim$ 13.

\subsubsection{DY\,Ori}
\citet{schmidt95} studied the pulsations of DY\,Ori and they showed that it is a 
photometric RV Tauri variable but of an unknown type, with a pulsation period of 60\,$\pm$\,8 days. The analysis was 
based on only 21 V-Band photometric observations from the General Catalogue of Variable Stars (GCVS).

One of the first optical spectral analyses of this object was presented by \citet{evans85}, after which it was classified as an RV Tauri 
spectroscopic type B based on the system of \citet{preston63}. \citet{gonzalez97a} carried out a detailed abundance 
analysis of field RV Tauri stars and DY\,Ori was found to have a highly depleted 
photosphere and very iron-poor with an [Fe/H] of -2.3\,$\pm$\,0.11\,dex. \citet{gonzalez97a} also estimated the effective temperature 
of DY\,Ori as T$_{\rm eff}$ = 6000\,K using spectra at a phase just after maximum light.

\subsubsection{EP\,Lyr}
EP\,Lyr was first found to be variable by \citet{schneller31}. Much interest in the object was gained in the early 60's with the work of \citet{wenzel61}
where they claimed that it is a $\beta$-Lyrae type eclipsing binary with a long period. Later investigations by
\citet{preston63} and \citet{wachmann68} proved this claim to be wrong. EP\,Lyr was classified as an RVb photometric type
by \citet{kholopov85}, under the suspicion that the mean brightness was varying on a longterm.
After the work of \citet{zsoldos95}, it became clear that EP\,Lyr is of an RVa type with a pulsation period of 83.46 days
and no remarkable long-term photometric variability.

A synthetic spectrum analysis of EP\,Lyr by \citet{wahlgren92} reported an effective temperature of 5250\,K, log $g$ of 1.5, [Fe/H] of $-$1.7\,dex.
\citet{gonzalez97a} carried out an abundance study of EP\,Lyr in which they better constrained the T$_{\rm eff}$, log $g$ and [Fe/H] based on the Fe-line (see Table~\ref{table:sampleselection}).
They also found that the star is very metal-poor, rich in oxygen and very depleted.
The effective temperature of EP\,Lyr shows large changes across the pulsation phase from 5500\,K up to even 7000\,K \citep{gonzalez97a}.
This results in a spectral type change from G5 to A4. Based on the spectroscopic classification of \citet{preston63}, EP\,Lyr is classified as an RVB spectroscopic type.

\subsubsection{HP\,Lyr}
One of the first variability studies of this object includes the work of \citet{erleksova71}. They showed that its O$-$C diagram contain 
many period instabilities with abrupt changes from 0.001P to 0.01P on a scale of 20 to 100 pulsation cycles.
HP\,Lyr has been termed as a long period variable in the literature due to its high pulsation fundamental period of 69.35 days and a formal period of 138.7 days \citep{Graczyk2002}.
\citet{Graczyk2002} also showed one important feature of this star; the pulsation period is variable with at least 1\% change in 20 years. 
On the other hand, it does not show any long-term variability in its mean V-mag, due to which it is classified as a photometric RVa type.

HP\,Lyr is of an RVB spectroscopic type based on the classification of \citet{preston63}. The star shows a few peculiarities in the RV Tauri class; 
it was reported as probably the hottest known RV-Tauri variable by \citet{Graczyk2002} with a mean effective temperature of 7700\,K.
A spectral analysis of HP\,Lyr by \citet{giridhar05} resulted in a T$_{\rm eff}$ of 6300\,K.

This object might be one where the dust-gas separation is most apparent.
Fig. 4 of \citet{giridhar05} shows a tight correlation between different chemical species with respect to solar abundance and the condensation temperature, T$\rm _C$.
Elements like C, N and O are close to solar abundance and high condensation temperature elements like Ti, Al and Sc are highly under-abundant.

\subsubsection{IRAS\,17038$-$4815}
Despite being a genuine RV Tauri star, IRAS\,17038$-$4815 has got little attention in the past decades. 
It has a high amplitude pulsation of $\sim$ 1.6 mag in the V-band and a photometric fundamental 
period of 38 days, with a formal period of 76 days \citep{maas03b}. \citet{kiss07} classifies it as an RVa photometric type based on the only found period
of 75.9 $\pm$ 1.9 days in the ASAS photometric time series.

The strong CN and CH bands in its spectrum makes it a genuine RVB spectroscopic type. \citet{maas05} derives an effective temperature
of 4750 K and a metallicity of [Fe/H] = $-$1.5\,dex, indicating a depleted atmosphere.

The preliminary orbit of IRAS\,17038$-$4815 was already published by \citet{vanwinckel07}. They found a pulsation period of 37.95 days which
was cleaned from the orbit to obtain an orbital period of 1381 $\pm$ 16 days and eccentricity of
0.56 $\pm$ 0.05.

\subsubsection{IRAS\,09144$-$4933}
A recent variability study of IRAS\,09144$-$4933 is presented in the work of \citet{kiss07}. They analysed the ASAS photometry and 
estimated a pulsation period of 93 $\pm$ 2 days. The fit to their pulsation data is not very clear and the ASAS 
data itself seem to have a lot of scatter. This object might be either too faint for pulsations to be detected by the 
ASAS instrument or a relatively low-amplitude pulsator. Note that this star is the faintest in our sample with a mean V-mag of $\sim$ 13.8.

IRAS\,09144$-$4933 is of a G0 spectral class with an effective temperature of 5750 K and a metallicity, [Fe/H] = $-$0.5\,dex. 
These parameters are derived assuming a plane parallel Local Thermodynamic Equilibrium (LTE) atmosphere \citep{maas05}. 
The elements with a high condensation temperature of $\sim$ 1500 K are either not or moderately depleted. 
\citet{maas05} also reports that most \textit{s}-process elements, except for Ba, Ti and Sc, are underabundant.

\subsubsection{TW\,Cam}
\citet{preston63} first reported a 
85 days photometric pulsation period using GCVS data. Later, \citet{zsoldos91} reported a fundamental period of 43 days together with a 86 days formal period. Over the last few decades, 
it became clear that these two periods are the dominant ones in the photometry of TW\,Cam. 
The star was reported as an RVa type by \citep{giridhar00}.

TW\,Cam was classified as an RVA spectroscopic type by \citet{preston63}. 
\citet{giridhar00} derived stellar model parameters for TW\,Cam from the Fe-lines and showed that it is mildly depleted with an [Fe/H] of $-$0.5\,dex. They also constrain an effective temperature of 4800 K from the Fe-line analyses. It therefore belongs to the spectral type G3.

\renewcommand{\tabcolsep}{0.2cm}
\begin{table*}[ht]
\tiny

 \centering                          % used for centering table
 \begin{tabular}{p{2.5cm} | l l p{0.8cm} p{1.0cm} p{0.8cm} p{0.9cm} p{0.6cm} p{0.8cm} p{1.5cm}}        % centered columns (4 columns)
 \hline\hline                 % inserts double horizontal lines
   Star  & RA & DEC & $V$ & S.Type & T$_{\rm eff}$ & Log \textit{g} & [Fe/H] & E($B$$-$$V$) & Ref.\\    % table heading 
	 &   (J2000) &  (J2000)   & (Mag) &  & (K) &  & (dex)  &  (Mag) &  \\
 \hline
   DY\,Ori 		& 06:06:14.9 	& $+$13:54:19.1  	& 11.29 		& G0I 	& 6000  & 1.5  & $-$2.0 	& 1.1	& b,c,e  \\
   EP\,Lyr 		& 19:18:19.5 	& $+$27:51:03.2   	& 10.40 		& A$-$G & 6200  & 1.2  & $-$1.5 	& 0.5	& c,e  \\
   HP\,Lyr  		& 19:21:39.1  	& $+$39:56:08.1   	& 10.40 		& A6 	& 6300  & 1.0  & $-$1.0 	& 0.1	& b  \\
   IRAS\,17038$-$4815  	& 17:07:36.6 	& $-$48:19:08.6 	& 10.55 		& G2p 	& 4750  & 0.5  & $-$1.5 	& 0.2	& d,e  \\
   IRAS\,09144$-$4933  	& 09:16:09.6 	& $-$49:46:12.2 	& 11.10 		& G0 	& 5750  & 0.5  & $-$0.3 	& 2.2	& d,e  \\
   TW\,Cam 		& 04:20:48.1  	& $+$57:26:26.5   	& 8.98 			& G3I 	& 4800  & 0.0  & $-$0.5 	& 0.4	& a,e \\
   \hline 
   \textbf{References:} \\
   \multicolumn{9}{l}{\text{a:} \footnotesize{\citet{giridhar00}}} \\
   \multicolumn{9}{l}{\text{b:} \footnotesize{\citet{giridhar05}}}  \\
   \multicolumn{9}{l}{\text{c:} \footnotesize{\citet{gonzalez97a}}} \\
   \multicolumn{9}{l}{\text{d:} \footnotesize{\citet{maas05}}} \\
    \multicolumn{9}{l}{\text{e:} \footnotesize{\citet{deruyter06}}} \\
 \end{tabular}
  \caption{Table showing our sample of Galactic RV Tauri stars. The second and third Columns display the coordinates. The 4$^{th}$ and
  5$^{th}$ Columns are the V-band magnitude and spectral type, respectively. The effective temperatures, surface gravity and metallicity obtained 
  from literature (see references displayed in Columns 6, 7 and 8, respectively. Column 9 lists the amount of extinction towards each star. 
  The references for each target are listed in Column 10.}            
   \label{table:sampleselection}     

 \end{table*}

\subsection{Photometric Data} \label{section:photometricdata}
We present an updated V-band photometric analysis based on the All Sky Automated Survey \citep[ASAS,][]{pojmanski02} to study the 
pulsations for all objects except for TW\,Cam and HP\,Lyr, for which data from the American Association of Variable Star Observers (AAVSO)\footnote{https://www.aavso.org/} were used.

ASAS is an ongoing long-term photometric monitoring that maps the sky south of $\delta$ = $+$28$^{\circ}$ consisting of
two observing sites namely, Las Campanas Observatory (Chile) and Haleakala (Maui).
Both are equipped with 2 wide-field instruments which can observe simultaneously in the V and I-band. 
The instrument is very well suited for sources with an utmost V-band brightness of 14 mag. To date, 
the ASAS catalogue consists of more than 267260 V-band frames that have imaged a total of around 
20 million stars brighter than 14 mag. Around 50000 of them are seen to be variable. 

We gathered ASAS data by querying the ASAS Catalogue of Variable Stars (ACVS)\footnote{http://www.astrouw.edu.pl/asas/?page=acvs} and the ASAS All Star Catalogue (AASC)\footnote{http://www.astrouw.edu.pl/asas/?page=aasc}. Each time series from ASAS is partitioned by its observed field.
Most of the data in our analysis were obtained using the ASAS-3 configuration with a wide-field (8.8$^{\circ}$\,$\times$\,8.8$^{\circ}$) CCD camera.

The critical PSF sampling is not reached by the instrument and therefore the pipeline data-reduction 
routine approximates the aperture photometry into 5 different aperture sized measurements which ranges 
from 2 pixels to 6 pixels. The ASAS guidelines recommend a preferred aperture size for every magnitude 
bin and we limited our photometric analysis with the A-grade data quality.

Table \ref{table:phot_data} outlines our photometric dataset. Column 2 represents the number of photometric data points, Column 3 gives the source of the data, Column 4 displays the ASAS aperture size used for each star and the last Column 
denotes the span of the time series.

\renewcommand{\tabcolsep}{0.11cm}
\begin{table}[ht]
\tiny
 \centering                        
 \begin{tabular}{p{2.5cm} | p{1.0cm} p{1.0cm} p{1.0cm} p{1.4cm}} 
 \hline\hline              
 Star 				& N$\rm _{phot}$ & Source 	  & Aperture	& $\triangle$T (days) \\
 \hline
 DY\,Ori 			& 	293  	 & ASAS	  	  & 	Mag 1   &  2544 	\\
 EP\,Lyr 			& 	44  	 & ASAS	  	  &     Mag 2	&  464 		\\
 HP\,Lyr 			& 	51  	 & AAVSO	  &	-	&  1075 	\\
 IRAS\,17038$-$4815 		& 	487	 & ASAS 	  &	Mag 2	&  3170		\\
 IRAS\,09144$-$4933 		& 	494	 & ASAS	  	  &	Mag 1	&  3623 	\\
 TW\,Cam 			& 	249	 & AAVSO	  &  	-	&  1237		\\
   \hline
 \end{tabular}
 \caption{Table showing the characteristics of the photometric dataset.}     
  \label{table:phot_data}   
 \end{table}

\subsection{Spectroscopic data} \label{section:HERMES_CORA}

High-resolution spectroscopic data for DY\,Ori, EP\,Lyr, HP\,Lyr and TW\,Cam were obtained using the HERMES spectrograph, 
mounted on the 1.2-m Mercator telescope at the Roque de los Muchachos Observatory, La Palma. HERMES is a fibre-fed spectrograph 
operating in a temperature-controlled environment to ensure good wavelength stability and excellent throughput. The fibre has an 
aperture of 2.5 arcsec on-sky and the high resolution is reached via a two-sliced image
slicer which is made to resemble a slit. It is built to have minimum light loss despite the high resolution. A star with a $V$-band magnitude of 10$^{th}$
has a S/N of 110 at 550\,nm in an hour exposure with an average seeing condition. The spectrograph obtains 55 orders with a resolution of R = 85\,000 
corresponding to 3.5\,km\,s$^{-1}$ and a high spectral coverage of 377$-$900\,nm in a single exposure. Once the observing night ends, the data are reduced 
through a dedicated pipeline tailored to ensure efficient data reduction \citep[see,][for details]{raskin11}.

The spectroscopic data for the targets in the southern hemisphere, IRAS\,17048$-$4815 and IRAS\,09144$-$4933, were obtained from the 1.2-m Swiss Leonhard Euler Telescope, located at La Silla observatory. 
The telescope is equipped with a fibre-fed echelle spectrograph (CORALIE), containing 68 orders with a broad wavelength range of 390\,nm~to 680\,nm. CORALIE can 
deliver a high resolving power of 50000 at 550\,nm \citep{queloz99}. 
The spectrograph is fed by 2 fibres; one of which is the science fibre and the other is a reference fibre. The science fibre has a diameter of 2 arcsec on-sky with a double scrambler device
for the input illumination to be more stable and the reference fibre is used for the calibration lamp.
The data are automatically reduced by a software that provides the measurement
of radial velocities via a cross-correlation method \citep[see,][for details]{baranne79,baranne96}.

Table \ref{table:spec_data} outlines our spectroscopic dataset; the original number of epochs per target (N$\rm _{original}$), the number of epochs with not good enough S/N in the spectra (N$\rm _{bad}$), the 
number of spectra related to shocks (N$\rm _{shocks}$, see section \ref{section:shocks}) and the final number of spectra that were considered good to be used for the orbital analysis 
(N$\rm _{used}$ = N$\rm _{original}$ $-$ N$\rm _{bad}$ $-$ N$\rm _{shocks}$). The span of the spectroscopic time series is denoted as $\triangle$T in the last Column of Table \ref{table:spec_data}. 
Note here that the minimum S/N in the spectra below which any radial velocity was treated as bad data due to poor weather conditions was set to 20.

\renewcommand{\tabcolsep}{0.11cm}
\begin{table}[ht]
\tiny
 \centering                        
 \begin{tabular}{p{2.5cm} | p{1.0cm} p{1.0cm} p{1.0cm} p{1.0cm} p{1.4cm}} 
 \hline\hline              
 Star 					& N$\rm _{original}$	& N$\rm _{bad}$ 	& N$\rm _{shocks}$ 	& N$\rm _{used}$	& $\triangle$T (days) \\
 \hline
 DY\,Ori 				& 79			& 4			& 4 			& 71			& 2306 		\\
 EP\,Lyr 				& 78			& 11 			& 9 			& 58			& 2272 		\\
 HP\,Lyr 				& 68			& 1			& 17			& 50			& 1572 		\\
 IRAS\,17038$-$4815 			& 98			& 5 			& 18 			& 75			& 1762 		\\
 IRAS\,09144$-$4933 			& 88			& 0			& 14 			& 74			& 3778 		\\
 TW\,Cam 				& 126			& 0 			& 39 			& 87			& 2297 		\\
   \hline
 \end{tabular}
 \caption{Table showing the characteristics of the spectroscopic dataset. See text.}     
  \label{table:spec_data}   
 \end{table}

\section{Photometric Analysis and the PLC relation} \label{section:photometrypulsations}
\subsection{Pulsation analysis}

The main aim of this research is to probe the binary status of our sample of RV Tauri stars with a disc. 
However, these are high-amplitude pulsators. In some cases (e.g, low-inclined orbital planes) the pulsations might even drown the orbital 
motions in the radial velocity amplitude. To better constrain the orbital parameters of the binary system, the radial velocity variations induced by the pulsations should be 
cleaned from the observed radial velocities. 

The original photometric data was first checked for any extreme outliers 
and these points were excluded by imposing a sigma clipping to any flux value beyond $\sim$ 4\,$\sigma$ 
from the mean. We then carried out a pulsation analysis in the photometric time series to obtain the dominant pulsation frequencies. 
We limit ourselves to a signal-to-noise (S/N) ratio of $\sim$\,4 in the analysis of the frequency spectrum.
This S/N is in principle the accepted critical value below which one cannot identify signal from noise \citep{breger93}.
The photometric pulsation periods are used to disentangle the pulsations from the original radial velocity data. However, the cycle-to-cycle variability in the pulsations also induces a signal in the radial velocities, which is hard to correct for.

\begin{figure}[h]
   \centering
   \includegraphics[width=8cm,height=6cm]{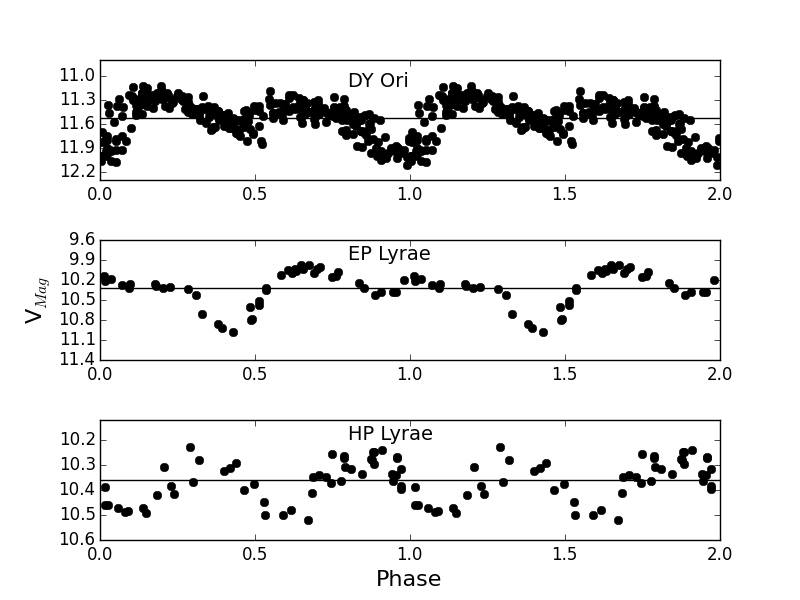}
   \caption{A photometric phase plot of DY\,Ori, EP\,Lyr and HP\,Lyr, folded on a formal period of 60.8 days, 83.2 days and 136.8 days, respectively.
   The horizontal solid black line is the mean V$\rm _{mag}$}.
     \label{figure:all_ASAS_pulsationsphase1}
\end{figure}

\begin{figure}   [h]
   \centering
   \includegraphics[width=8cm,height=6cm]{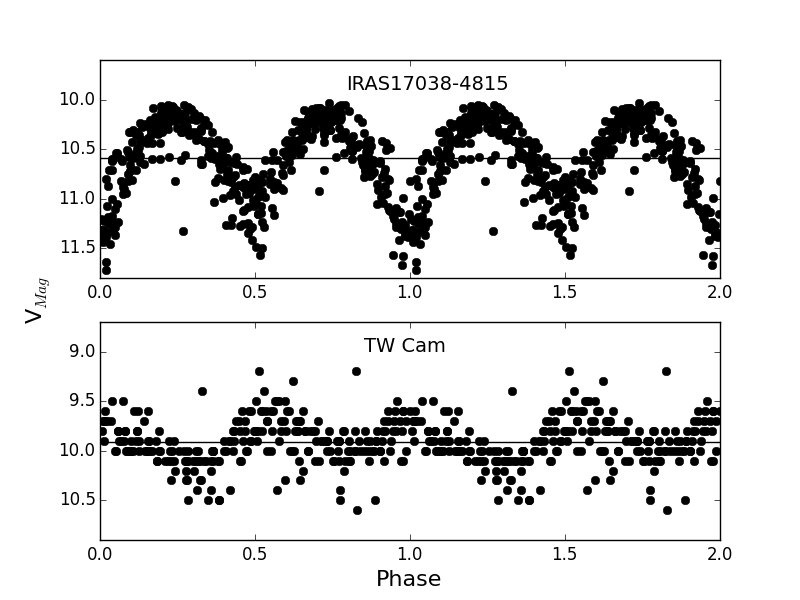}
   \caption{A photometric phase plot of IRAS\,17038$-$4815 and TW\,Cam folded on a formal period of 75.9 days and 86.1 days, respectively.
   The horizontal solid black line is the mean V$\rm _{mag}$}.
     \label{figure:all_ASAS_pulsationsphase2}
\end{figure}

The pre-whitening procedure involves finding the dominant peak frequency using the
Lomb-Scargle method \citep{scargle82}. The fit parameters associated to the dominant frequency is fit using a sinusoidal function
and then subtracted from the original data to find the next significant peak frequency in the residuals. This process is repeated 
until the S/N stop-criterion is reached. The pulsation frequencies obtained above the S/N criterion are listed in Table~\ref{table:Oparameters_puls}. 
We used the fundamental period (P$_0$) obtained after the pre-whitening process, to fold the data on the formal period (P$_1$ = 2 $\times$ P$_0$). The phased
pulsation periods are shown in Figs. \ref{figure:all_ASAS_pulsationsphase1} and \ref{figure:all_ASAS_pulsationsphase2}.

The pulsation periods derived for DY\,Ori, EP\,Lyr, HP\,Lyr and TW\,Cam are consistent with what has been presented in 
the literature (see references in Table~\ref{table:Oparameters_puls}). In the case of EP\,Lyr, we also consider the frequency peak with S/N = 3.8, 
corresponding to a pulsation period of 41.6 days, as a real signal (as quoted in the literature).

The RVb nature of IRAS\,17038$-$4815, IRAS\,09144$-$4933 and TW\,Cam are 
confirmed by the long term photometric variation in the light curves on the time-scale of the orbital period 
(see italicised numbers in Column 3 of Table \ref{table:Oparameters_puls}). No long-term photometric variability is seen in the time series 
of DY\,Ori extending over $\sim$ 2500 days ($\sim$\,2\,$\times$ P$_{orb}$). We therefore classify it as an RVa photometric type.

\renewcommand{\tabcolsep}{0.1cm}
\begin{table*}[ht]
  \begin{small}
 \centering                        
 \begin{tabular}{p{2.5cm} | p{1.8cm} | p{3.2cm} p{2.3cm} p{3.0cm} |p{1.4cm}}    
 \hline\hline                
					& \centering \bfseries Spectroscopy 	&						&   	\centering \bfseries Photometry &			&   \\ \hline
    \centering Star 	  		&  \centering P$_{Puls.}$ 		& \centering P$_{Puls.}$ 			& \centering S/N$_{Puls.}$		& \centering Amplitude	& Phot. type    \\  
					&  \centering (days) 			& \centering (days) 				&					& \centering (mag)	&       \\
 \hline                     
   DY\,Ori     	  			&    30      	 			& 30.4, 60.9 $^2$         			& 7.7, 6.7  				&0.23, 0.11		& RVa$^a$    \\                    
   EP\,Lyr         			&    42, 83  	 			& 41.6, 83.8$^2$      				&  3.8, 4.6 				&0.25, 0.22		& RVa$^b$    \\
   HP\,Lyr         			&    68      	  			& 68.4$^1$              			&  4.8					&0.11			& RVa$^c$     \\
   IRAS\,17038$-$4815 			&    38		 			& 37.9, 19.0, 69.3, \textit{1356.2}$^{2*}$ 	& 16.0, 5.3, 4.8, 4.9  			&0.44, 0.09, 0.08, 0.08	& RVb$^a$      \\
   IRAS\,09144$-$4933 			&    50     				& \textit{1758.0}$^*$, 85.4, 50.3$^2$ 		& 6.0, 4.6, 4.0				&0.12, 0.07, 0.07	& RVb$^a$	       \\ 
   TW\,Cam         			&    43      	    			& 43.5, \textit{671.0}$^{1*}$ 			& 7.2, 5.9				&0.08, 0.07		& RVb$^a$      \\

 \hline
 \end{tabular}
 \caption{Table listing the pulsation periods derived from the photometric and spectroscopic time series. Column 2 shows the pulsation periods in the radial 
 velocities with peaks in the Lomb-Scargle periodogram above a S/N of 4. Column 3 shows the pulsation periods found in the photometry. We indicate in Column 4 
 the S/N in the Lomb-Scargle periodogram associated with each pulsation period in the photometry. The amplitudes of the pulsations are shown in Column 5.}  
  \label{table:Oparameters_puls}     
 \tablefoot{\\ \tablefoottext{1}{Data from AAVSO} \\ \tablefoottext{2}{Data from ASAS} \\ \tablefoottext{a}{This work} \\ \tablefoottext{b}{\citet{zsoldos95} \\ \tablefoottext{c}{\citet{Graczyk2002} \\ \tablefoottext{*}{The italicised number in the 3$^{rd}$ col. are the spectroscopic orbital periods found in the photometry, which confirms their RVb nature.}}}}

    \end{small}
 \end{table*}

\begin{figure*}
   \centering
   \includegraphics[width=14cm,height=16cm]{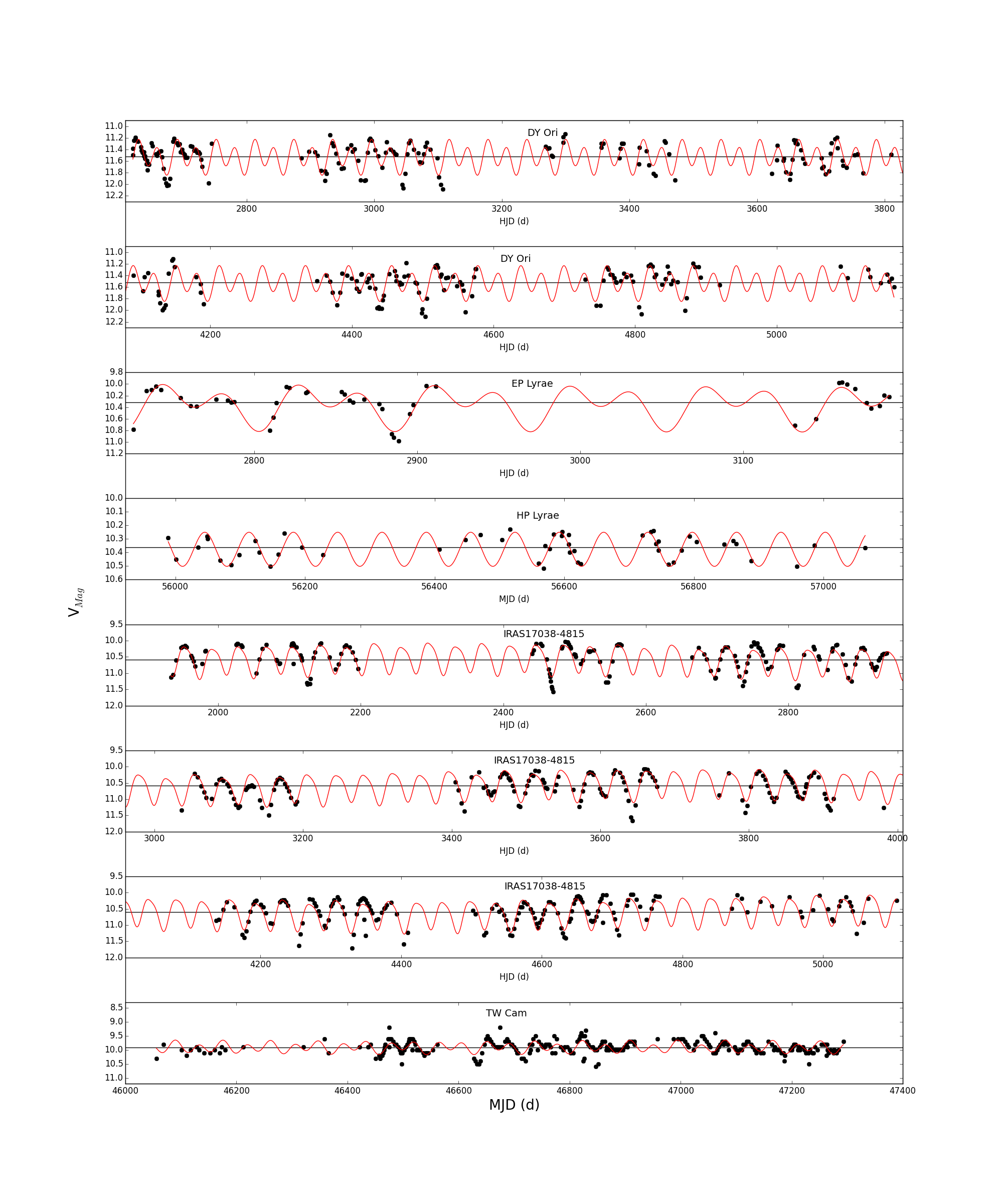} 
   \caption{time series of the pulsations using ASAS and AAVSO photometric data. Here we mainly display their pulsational nature and not the long-term variability.
   The solid red is the pulsation model fit to the photometric time series and the horizontal solid black line is the mean V$\rm _{mag}$}.
     \label{figure:all_ASAS_pulsationstimes2}
\end{figure*}

Fig.~\ref{figure:all_ASAS_pulsationstimes2} displays the time series of the 
photometric pulsations. Note that we did not include the pulsation time series of IRAS\,09144$-$4933 in
Fig.~\ref{figure:all_ASAS_pulsationstimes2}, because the main period detected in the light curves is the orbital period (P = 1758 days, see Table \ref{table:Oparameters_puls}). Two putative pulsation periods of 85 and 
50 days are detected only slightly above the S/N criterion (see Table~\ref{table:Oparameters_puls}), but the scatter in the data 
is too large to obtain a good fit to the pulsations. 

 %/home/rajeev/hermesRun/pulsations/ASASPhotometry/allplots_time.py

\subsection{Spectral Energy Distribution} \label{section:SED}
We model the SEDs of the stars using photometric data available from different catalogues listed in Tabs.~\ref{table:dyori_sedphot} to~\ref{table:twcam_sedphot} (see,  Appendix \ref{appendix:AppendixB}).
The fit itself is done using a parameter-grid search to obtain the optimised Kurucz model \citep{kurucz79} parameters using a $\chi^2$ minimisation method, see \citet{degroote2011} for details. 
We used initial model atmosphere parameters, T$_{\rm eff}$ and [Fe/H], based on the values found in the literature, listed in Table~\ref{table:sampleselection}. 
We note that the [Fe/H] was kept fixed in the SED fit, to the values found in the literature.
Since the T$_{\rm eff}$ changes significantly during the pulsation cycle, we account for the T$_{\rm eff}$  range during the pulsation cycle by allowing the effective temperature to vary by 10\% of its value.
An example of the fitted SED for DY\,Ori is shown in Fig. \ref{figure:SEDDYOri} and the 
rest of the SEDs are displayed in Appendix \ref{appendix:AppendixA}.

\begin{figure}[h]
   \centering
   \includegraphics[width=8cm,height=6cm]{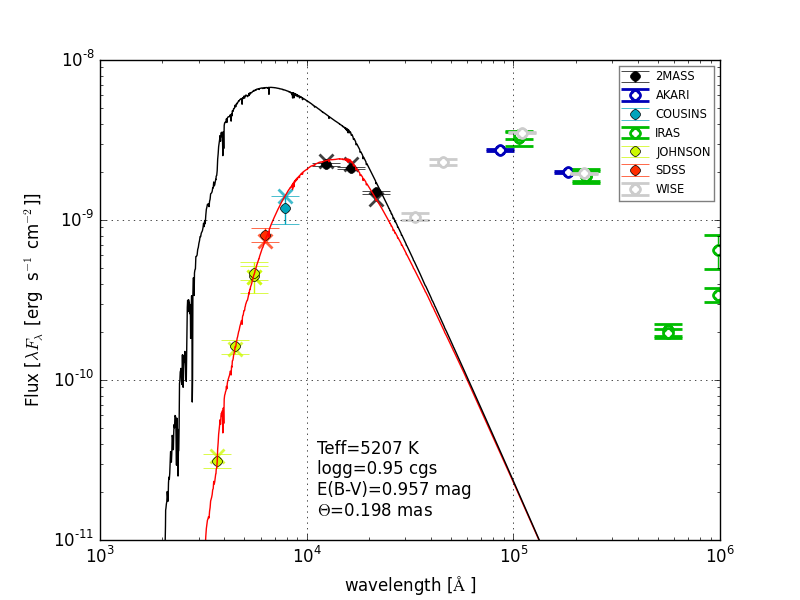} 
   \caption{The SED of DY\,Ori. The red model is the stellar photosphere fitted using a Kurucz model and the black 
   line is the dereddened model.}
     \label{figure:SEDDYOri}
\end{figure} %/home/rajeev/hermesRun/pulsations/ASASPhotometry/allplots_time.py

From the SEDs we obtain the total extinction E($B$$-$$V$)  towards the star and the results are illustrated in Table \ref{table:distLTeff}. We assume in this process
that the colour excess is due to interstellar and circumstellar reddening. We also assume that the circumstellar component follows the same extinction law.
The errors in E($B$$-$$V$)  are computed using a Monte-Carlo method on the photometric SED data whereby a random number generator was
used to scramble the original SED datapoints within one standard deviation creating 100000 number of randomized SED fit of the same structure. The errors 
were then computed by taking the standard deviation of the parameters linked to each SED fit.

The main challenge in building SEDs for strong pulsators like RV Tauri stars is that the photometric flux changes
considerably from one phase to the other. The way to compensate for that would be to obtain similarly phased data. As these data are not available, we restricted our fit 
to the photometry we retrieved which includes all available phases.

\subsection{The period-luminosity-colour (PLC) relationship and distances} \label{section:PL}
One of the main advantages that pulsating type II cepheids offer is that they allow us to constrain
the distances and hence luminosities via a Period-Luminosity (P-L) relation. The magellanic clouds have proven to be particularly 
useful in the calibration of the P-L relation for population II cepheids. This is mainly because they 
are close enough for reasonably bright apparent magnitudes of the stars and distant enough to approximate their distance as the same.
Some of the P-L relationship studies of the population II cepheids in the Galactic globular clusters, small magellanic cloud (SMC) 
and large magellanic cloud (LMC) are included in the work of \citet{nemec94,alcock98,ripepi2015}.

Since the luminosity of a star is dependent on two variables; the radius and effective temperature, we expect the luminosity of
a type II cepheid to be better constrained using a colour and period relation. Indeed, this proves to be the case since
the PLC relationship is generally tighter than the P-L relation. 
We calibrate the PLC relation in the LMC and apply it to our Galactic stars, assuming there is no 
metallicity dependency of the relation. 

The amount of extinction towards a star plays an important role in obtaining a well calibrated PLC relation. 
To circumvent and minimise reddening effects, we do not use the $V$ and $I$ band apparent magnitudes independently but instead we use
the Wesenheit index \citep{ngeow05} which is a combination of both. The Wesenheit index is reddening-free and is defined as,

\begin{equation}
 W_{I} = V - 2.55(V - I) 
\end{equation}

The reddening-free Wesenheit index can also be 
written in terms of the intrinsic values of the same variables \citep{vandenbergh68,madore76,madore82},
\begin{equation}
 W_{I} = V - 2.55(V - I) = V_{0} - 2.55(V - I)_{0}
\end{equation}

The PLC relation is constructed based on publicly available data\footnote{ftp://ftp.astrouw.edu.pl/ogle/ogle3/OIII-CVS/lmc/t2cep/} \citep{soszynsky2008a}. 
These are data from the OGLE-III observations carried out using the 1.3\,m Warsaw 
telescope at Las Campanas Observatory in Chile. Stars with either a missing I-band photometry or Fourier coefficient ($\phi_{31}$) were not included. 
This lead to a final list of 187 LMC type II Cepheids.
We decided to recalibrate the PLC relation using a broader log\,(P) range of
0.001 $\leqslant$ Log\,(P) $\leqslant$ 1.83, available for a bigger sample of type II Cepheids compared to the smaller sample 
and a narrower log\,(P) range of 0.9 $\leqslant$ log\,(P) $\leqslant$ 1.75, studied by~\citet{alcock98}.

\begin{figure}[h]
   \centering
   \includegraphics[width=9cm,height=12cm]{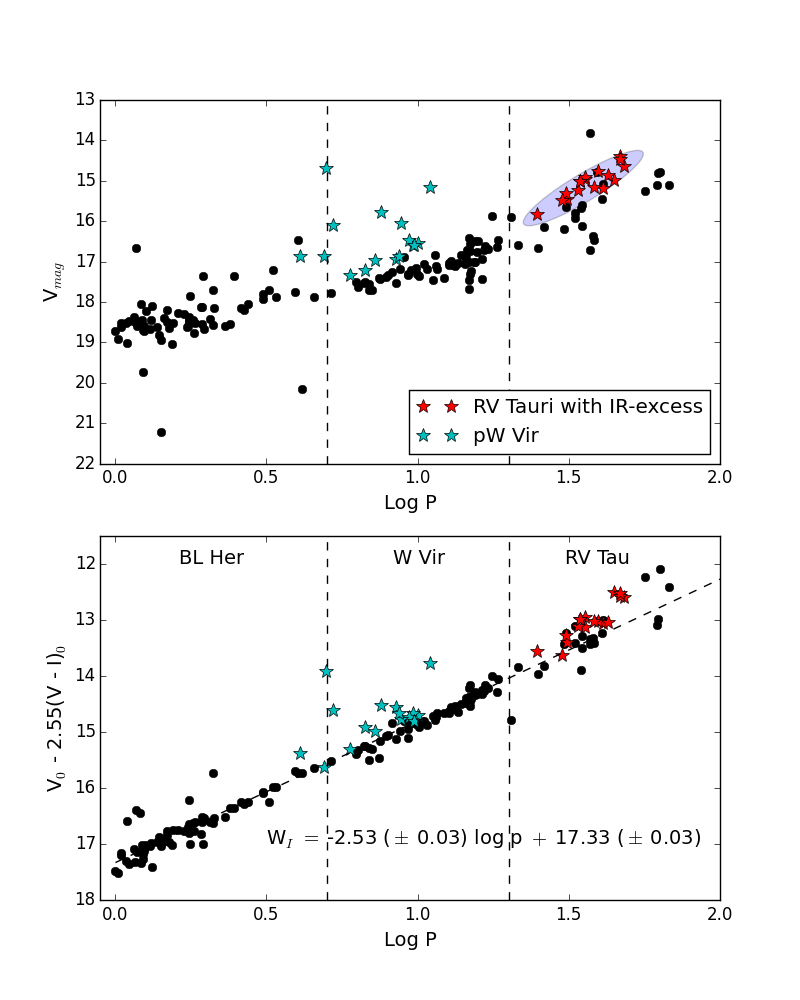}
   \caption{\textit{Top:} Apparent V-band magnitude (m$_v$). \textit{Bottom:} Reddening-free Wesenheit magnitudes for population II cepheids in the LMC. }
     \label{figure:PL_galactic_lmc}
\end{figure}

We used the mean I-band, mean V-band and periods with their associated uncertainties to obtain the quantity, $V_{0} - 2.55(V - I)_{0}$ for each log\,(P) value. 
A significant reduction in scatter in the Wesenheit plane is seen (Fig.~\ref{figure:PL_galactic_lmc}). This advocates the fact that the stars in the sample are
affected by a considerable amount of reddening. The few outliers in the Wesenheit plane in the W Vir regime is due to the peculiar\,W\,Vir stars (pW\,Vir; marked with Cyan star symbols). Other outliers in the 
BL\,Her region could possibly be attributed to either eclipsing type II Cepheids or anomalous type II Cepheids which are different from the W\,Vir, pW\,Vir, BL\,Her and RV Tauri stars \citep[for more details, see,][]{soszynsky2008a}.

Our results for the BL\,Her and W\,Vir types are in accordance to the results of \citet{matsunaga09} which show that these 2 types of population II cepheids are colinear in the P-L plane.
We notice a relatively larger scatter in the quantity $V_{0} - 2.55(V - I)_{0}$  and an upward ``bending'' of the 
relation for the longer periods (> 20 days), in the RV Tauri stars' regime. This can be attributed to extinction by
circumstellar dust. It is more apparent in the J and K$_s$-band, as these bands are more contaminated by thermal emission of the hot dust \citep[e.g,][]{matsunaga09,ripepi2015}. 
To optimise our PLC relationship, we excluded the objects with circumstellar dust in a linear regression analysis. The red star symbols in Fig.~\ref{figure:PL_galactic_lmc}
identifies the highly reddened RV Tauri stars in the LMC.

We made use of the extinction law (A$_{I}$/A$_V$) by \citet{cardelli89}, to obtain the intrinsic ($V$$-$$I$)$_0$ from the individual ($V$$-$$I$) values
of each star in the LMC.

We assumed a mean reddening towards the LMC of E($V$$-$$I$) = 0.09 $\pm$ 0.07 \citep{haschke2011}, consistent with the mean E($V$$-$$I$) of 
population II Cepheids in the LMC by \citet{ripepi2015}. A linear regression in Log (P) against the quantity $V$$_0$ $-$ ($V$$-$$I$)$_0$, yields into the following PLC relation for the intrinsic apparent magnitude ($V$$_0$) of the population II cepheids,
\begin{equation} \label{equation:vo}
V_{0} = -2.53\ (\pm 0.03)\ Log P + 17.33\ (\pm 0.03) + 2.55\mean{(V - I)_{0}}
\end{equation}

Adopting a distance modulus to the LMC of 18.5 \citep{walker2012}, we compute the absolute magnitudes of the population II cepheids in the LMC as,

\begin{equation} \label{equation:MVLMC}
M_{V} = -2.53\ (\pm 0.03)\ Log P - 1.17\ (\pm 0.03) + 2.55\mean{(V - I)_{0}}
\end{equation}

Based on 30 LMC type II Cepheids, \citet{alcock98} derives a steeper slope of $-$2.95 $\pm$ 0.12 and a higher zero-point offset of 17.89 $\pm$ 0.2 in Equation \ref{equation:vo}.
The PLC relation for the population II cepheids in the LMC is shown in Fig.~\ref{figure:PL_galactic_lmc}.

It is straightforward to derive the distances to our Galactic targets using the absolute magnitudes from the PLC relationship.
However, one still need to take into account the amount of extinction, E($B$$-$$V$) towards the stars. The individual E($B$$-$$V$) values of our Galactic sample were obtained by computing a SED Kurucz model and apply an amount of reddening to match the observed values. We used a value for the total to selective Galactic extinction, 
R$_v$ of 3.1 \citep{Weingartner2001} to obtain the amount of extinction A$_v$ towards each star in our sample and hence the reddening-corrected distances. The luminosities were then computed using the distances and the total integrated flux below the dereddened SED model. The results are outlined in Table~\ref{table:distLTeff}. 

The error in the distances and hence luminosities is a combined error in the absolute magnitude obtained from the PLC and the uncertainty in the amount of reddening.

\renewcommand{\tabcolsep}{0.2cm}
\begin{table}[ht]
\tiny
 \centering                          
 \begin{tabular}{l|ccccc}        
 \hline\hline                 
Star 		 & P$_{puls.}$ & E($B$$-$$V$)  & T$_{\rm eff}$ & Distance & Luminosity\\
		 & (d) & (mag) & (K) & (kpc) & (L$_{\odot}$)\\
\hline
DY\,Ori 		 & 30.4  & 0.9 $\pm$ 0.1 & 5500 &  2.1 $\pm$ 0.21	& 1350   $\pm$ 250 \\
EP\,Lyr 		 & 41.6  & 0.4 $\pm$ 0.2 & 6200 &  3.0  $\pm$ 0.30	& 1700  $\pm$ 300 \\
HP\,Lyr 		 & 68.4  & 0.1 $\pm$ 0.1 & 5900 &  6.7  $\pm$ 0.38 	& 3900  $\pm$ 400  \\
IRAS\,17038 	 & 37.9  & 0.3 $\pm$ 0.1 & 5000 &  3.8  $\pm$ 0.46	& 3000   $\pm$ 700 \\
IRAS\,09144 	 & 50.3  & 1.4 $\pm$ 0.2 & 5500 &  2.5  $\pm$ 0.36	& 2800  $\pm$ 800 \\
TW\,Cam 		 & 43.1  & 0.3 $\pm$ 0.1 & 4700 &  2.7 $\pm$ 0.26 	& 3000  $\pm$ 600 \\
 \hline
 \end{tabular}
 \caption{Column 2 shows the fundamental pulsation period (P$_0$). Column 3 is the amount of extinction towards each object obtained from the reddened SED model.
 The temperatures used for the SED fitting are shown in Column 4. The distances and luminosities are shown in the last 2 Columns, respectively.}
  \label{table:distLTeff}    
 \end{table}

\section{Spectroscopic results} \label{section:spectroscopy}

\subsection{Radial Velocities}

The individual radial velocities are obtained by cross-correlating the reduced spectra with a predefined software mask adapted for a given object type. 
A typical mask contains a large set of distinct spectral lines for a given stellar spectral type. 

We made use of a F0 and G2 mask for HP\,Lyr and TW\,Cam, respectively. Depending on
S/N quality criterion, a typical number of lines used is 1200 for both masks and both stars.
We only used lines obtained in spectral orders between 54 and 74 (477-655\,nm) which are typically the spectral orders with the highest S/N
and without strong telluric contamination.

Being very depleted, the spectra of DY\,Ori and EP\,Lyr do not show many photospheric lines. These are the 2 stars in the sample for which a mask adapted to their photospheric anomalies, had to be constructed.
This is done by plotting a spectrum with high overall photon count and identifying the main stellar lines present. 
The mask is then a linelist containing the rest wavelengths of the stellar lines identified. The mask for DY\,Ori and EP\,Lyr
contain 72 stellar lines covering a wavelength range of 434.2\,nm - 746.8\,nm. We outline here that a spectrum with an overall good S/N
does not necessarily point to a good S/N in the cross-correlation function (CCF). The main factor describing the S/N in the CCF is how optimised the cross-correlation mask is.

The radial velocities for IRAS\,17038$-$4815 and IRAS\,09144$-$4933 from CORALIE are derived in the same way as the HERMES data using a mask suited for a star of spectral type G2.

Each radial velocity point itself is computed by fitting a Gaussian to the cross-correlation function and computing the mean of the fit. 
The error in each radial velocity data point is then one standard deviation of the Gaussian fit. The data points with a bad S/N in the CCF even after using an optimised mask, were discarded, 
as these are most likely to be related to spectra taken in poor weather conditions. The S/N criterion in the CCF below which the data was treated as ``bad'' was different from star to star.
\subsubsection{Shocks} \label{section:shocks}
The atmospheres of RV Tauri stars are known to show differential motion in their line forming layers. Thus, it is possible at certain phases that
2 waves are moving towards each other, creating shock regions. A study by \citet{fadeyev84} showed that indeed in very luminous stars, pulsations exist in the form of standing waves only 
in the innermost layers while in the outer layers the waves gets transformed into running waves. These shocks leave an observational signature in the spectrum of the star
in the form of distorted stellar line profile. A good example tracing the impact moment of the 2 waves is the appearance of the He\,I-5856 line as in the case of RU Cen \citep[e.g,][]{maas02}.
\citet{baird82} also mentions the apparition of the He\,I-5856 as emission in the spectrum of AC\,Her 
a certain phases in the pulsation. He\,I emission line can only be produced by a violent event, i.e a shock in the atmosphere of RU\,Cen. 

One very prominent signature of the shock is also seen in the H$_{\alpha}$ profile at certain phases. \citet{pollard97} showed an increased emission at given phases consistent with 
the passage of 2 shocks waves through the photosphere.

\begin{figure}[h]
   \centering
   \includegraphics[width=8cm,height=6cm]{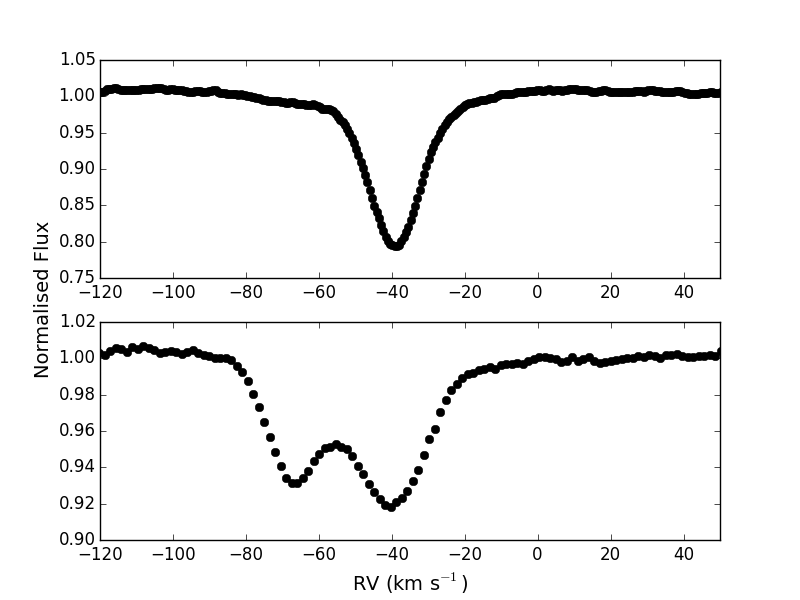}
   \caption{\textit{Top:} An example of a CCF profile unaffected by shock for TW\,Cam. \textit{Bottom:} The mean CCF profile of the cross-correlated stellar lines 
   in the spectrum with a double peak; a blue shifted and a redshifted peak indicative of 2 layers moving towards each other. }
     \label{figure:ccf_all}
\end{figure}

The CCF profile is the mean profile of the stellar lines cross-correlated with the mask. Therefore these shock events are seen as distortion 
in the mean CCF profile at certain phases in the pulsation cycle \citep[e.g,][]{gillet90}. A double peak is seen when the differential motion has a significant velocity difference. 
An example of such a situation is shown in Fig.~\ref{figure:ccf_all}. 

The velocities related to these distorted CCF profiles are identified and eliminated in our orbital analysis.
They are still plotted as ``zero-velocity-red-dots'' indicating at which phases in the pulsation cycle they occur (see mid-panel of Figs. \ref{figure:DYOriorbit} to \ref{figure:IRAS17038orbit}).

\subsection{Pulsation cleaned orbits} \label{section:cleaning}

To obtain the uncontaminated orbital velocities, we first fit the original radial velocities with a Keplerian orbital model using a non-linear least-square method.
We obtain the starting parameters of the fit by a minimalisation method. An eccentric orbital model (model 1) is constructed using the orbital period found in the periodogram.
We then compute a second eccentric model (model 2) after a least square minimalisation.
After each non-linear least-square fit, the variance reduction number is computed to compare model 1 and model 2.
The iteration continues until there is no significant change between 2 subsequent variance reduction values.

The residuals to the fit are then pre-whitened to find other periodicities which now trace the pulsations. The pulsation 
periods found above a S/N of 4 are cross-checked with the ones found in the photometry. If they match, the residuals are then fit with the formal period
and 2 overtones for all except HP\,Lyr and IRAS\,17038$-$4815 where 1 overtone was used. The pulsation fit is then subtracted from the original radial velocity 
data to obtain a pulsation-cleaned orbit. The cleaned orbit was then fit with a Keplerian model to extract accurate orbital parameters.
This procedure was repeated for all the stars in our sample.

The errors in the orbital parameters were computed using a Monte-Carlo method on the radial velocity data points. A random number generator was
used to scramble the datapoints with a mean value corresponding to the radial velocity value and a scaled standard deviation to match a fit with a $\chi^2$ value of $\sim$ 1.
This was done because the purely intrinsic uncertainties in the individual radial velocities were too small compared to the scatter in the data.
1000 randomized Keplerian fit of the same structure were obtained for each star and errors were then computed by taking the standard deviation of the parameters linked to each Keplerian fit.

\begin{figure}[h]
   \centering
   \includegraphics[width=8cm,height=6cm]{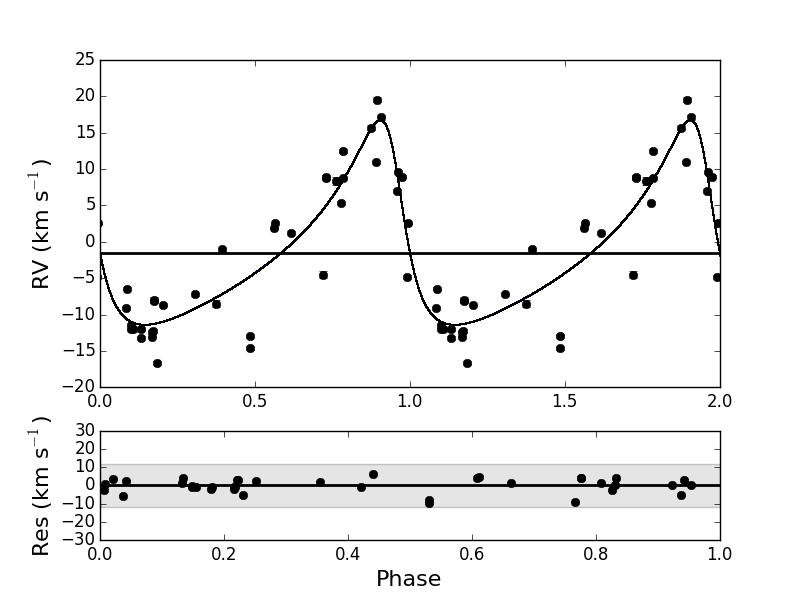}
   \includegraphics[width=8cm,height=6cm]{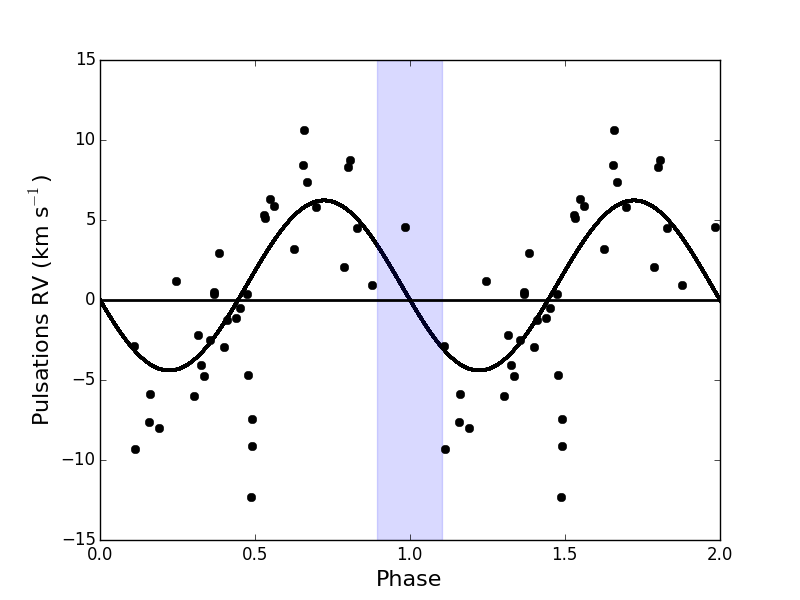}
   \includegraphics[width=8cm,height=6cm]{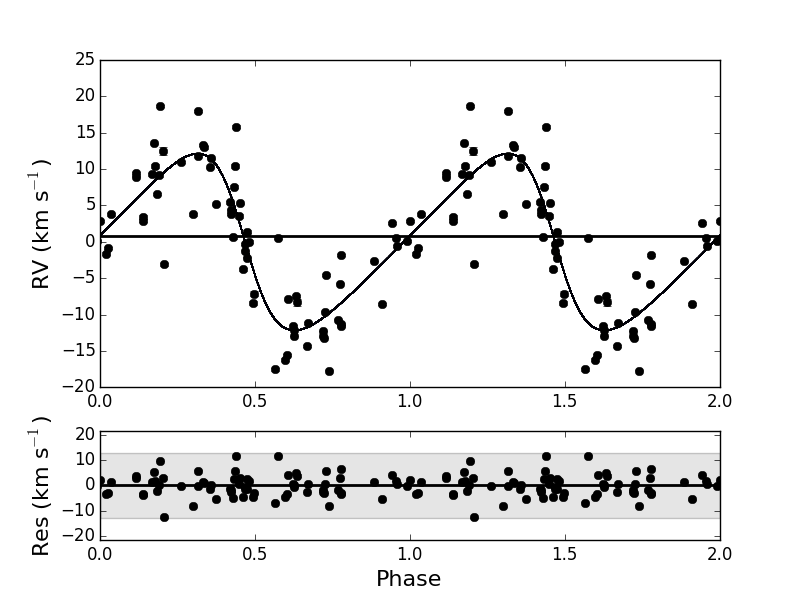}
   \caption{\textit{Top:} Original radial velocity data of DY\,Ori fitted with a keplerian orbit folded
   on an orbital period of 1244 d and eccentricity of 0.3. \textit{Middle:} Pulsations in the radial velocity folded on a period of
   60.8 d. The red square symbols and the purple shaded region displays the position in the phase plot where the shocks occur.
    \textit{Bottom:} Pulsation cleaned orbit folded on a period of 1230 d with an eccentricity of 0.2.
     }
     \label{figure:DYOriorbit}
\end{figure}

\begin{figure}[h]
   \centering
   \includegraphics[width=8cm,height=6cm]{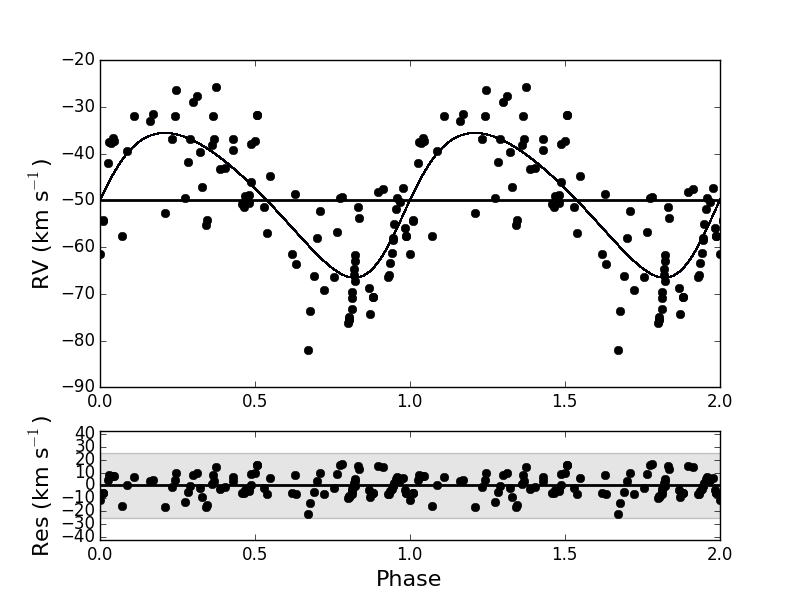}
   \includegraphics[width=8cm,height=6cm]{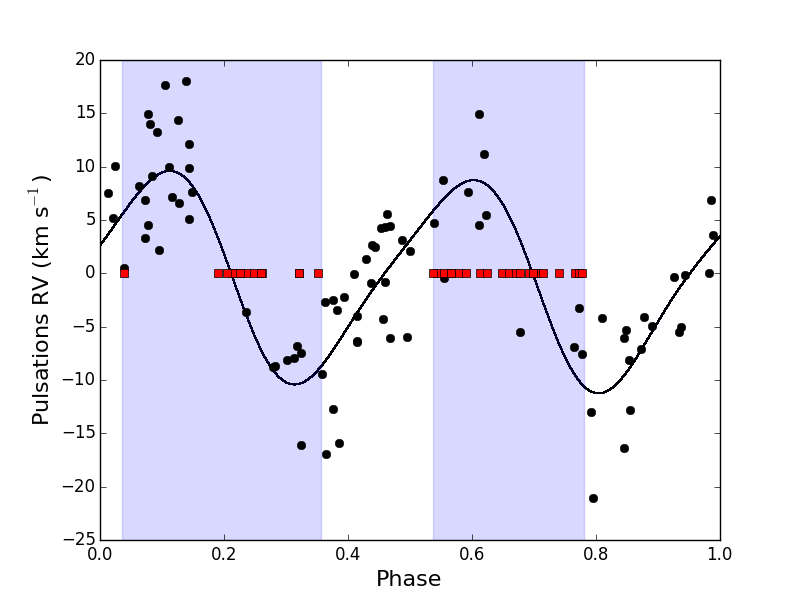}
   \includegraphics[width=8cm,height=6cm]{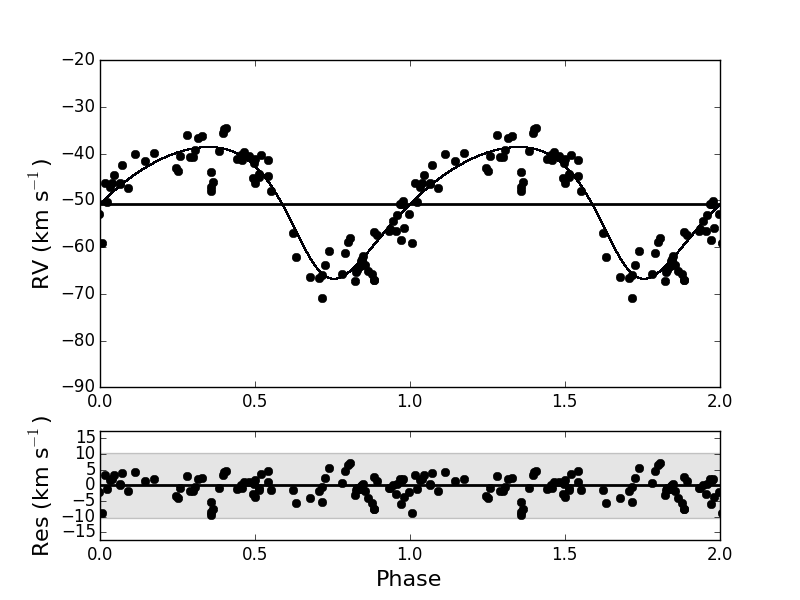}
   \caption{\textit{Top:} Original radial velocity data of TW\,Cam fitted with a keplerian orbit folded
   on an orbital period of 655 d and eccentricity of 0.3. \textit{Middle:} Pulsations in the radial velocity folded on a period of
   86.4 d. The red square symbols and the purple shaded region displays the position in the phase plot where the shocks occur.
    \textit{Bottom:} Pulsation cleaned orbit folded on a period of 654 d with an eccentricity of 0.2.
     }
     \label{figure:TWCamorbit}
\end{figure}

\begin{figure}[h]
   \centering
   \includegraphics[width=8cm,height=6cm]{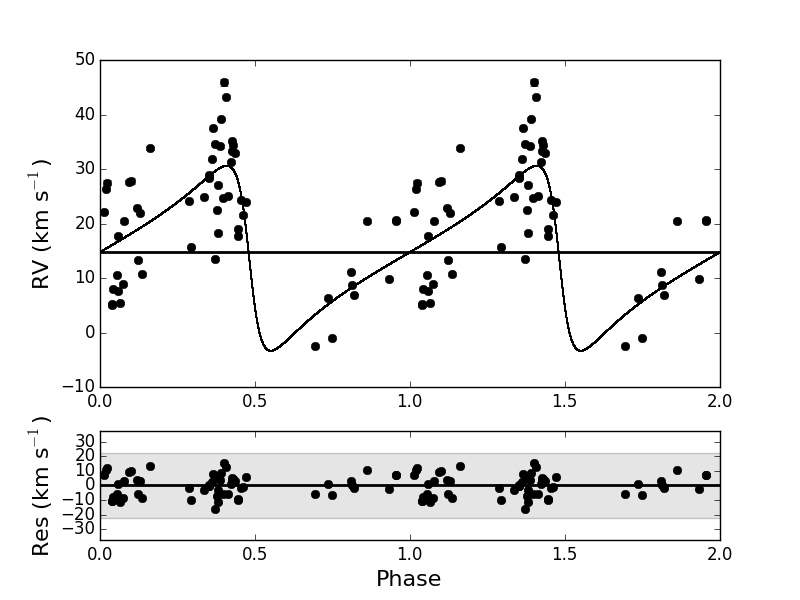}
   \includegraphics[width=8cm,height=6cm]{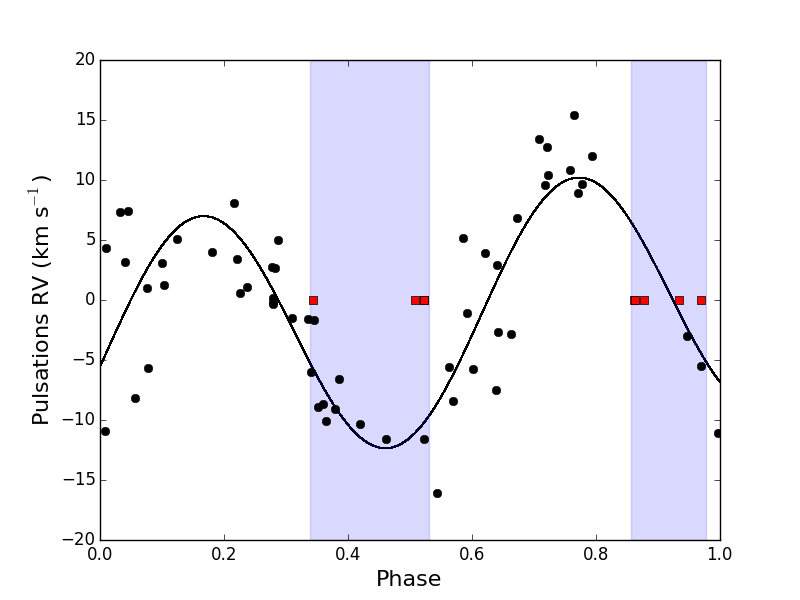}
   \includegraphics[width=8cm,height=6cm]{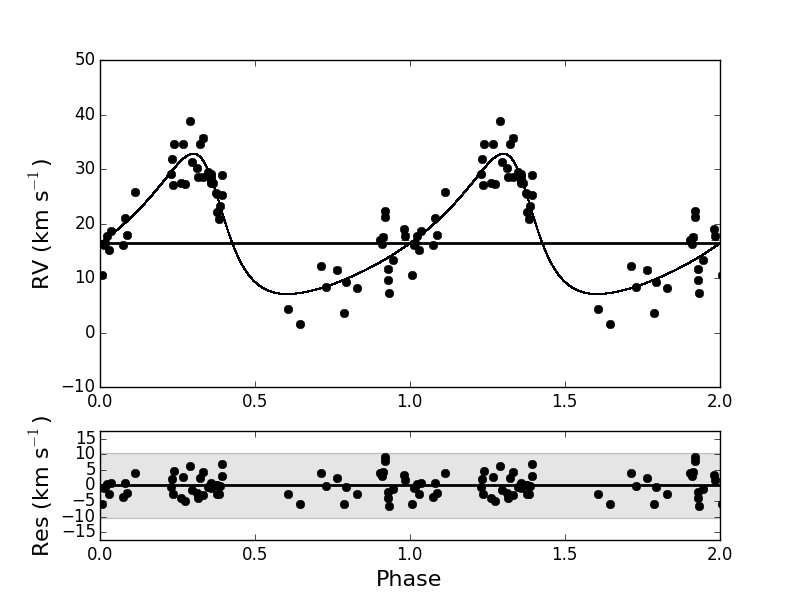}
   \caption{\textit{Top:} Original radial velocity data of EP\,Lyr fitted with a keplerian orbit folded
   on an orbital period of 1080 d and eccentricity of 0.6. \textit{Middle:} Pulsations in the radial velocity folded on a period of
   83.2 d. The red square symbols and the purple shaded region displays the position in the phase plot where the shocks occur.
    \textit{Bottom:} Pulsation cleaned orbit folded on a period of 1066 d with an eccentricity of 0.6.
     }
     \label{figure:EPLyrorbit}
\end{figure}

\begin{figure}[h]
   \centering
   \includegraphics[width=8cm,height=6cm]{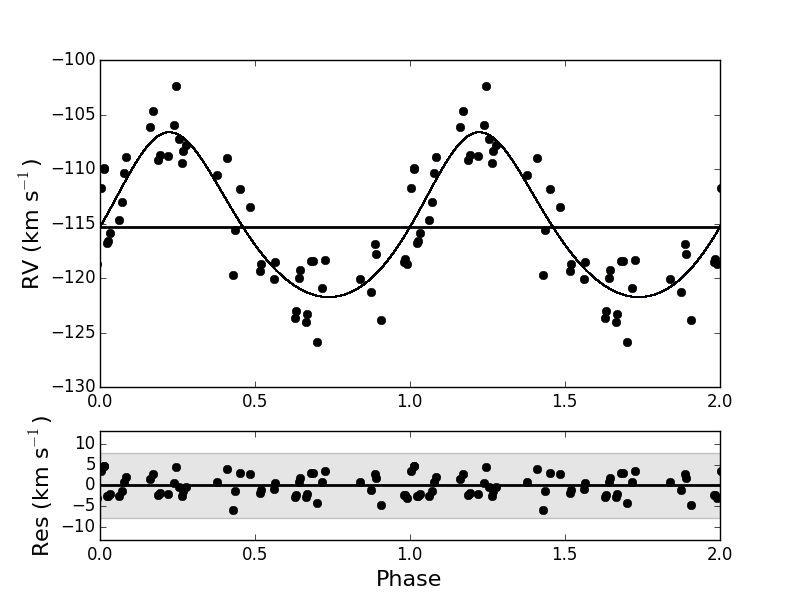}
   \includegraphics[width=8cm,height=6cm]{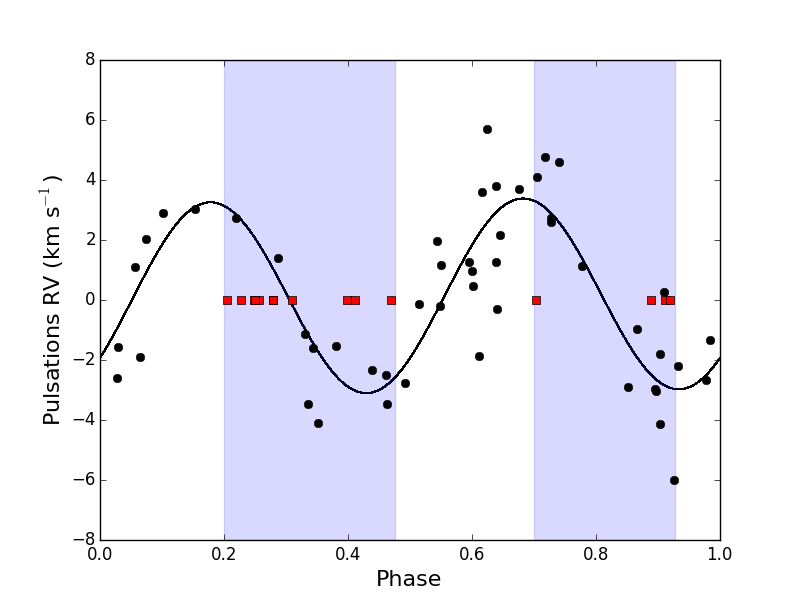}
   \includegraphics[width=8cm,height=6cm]{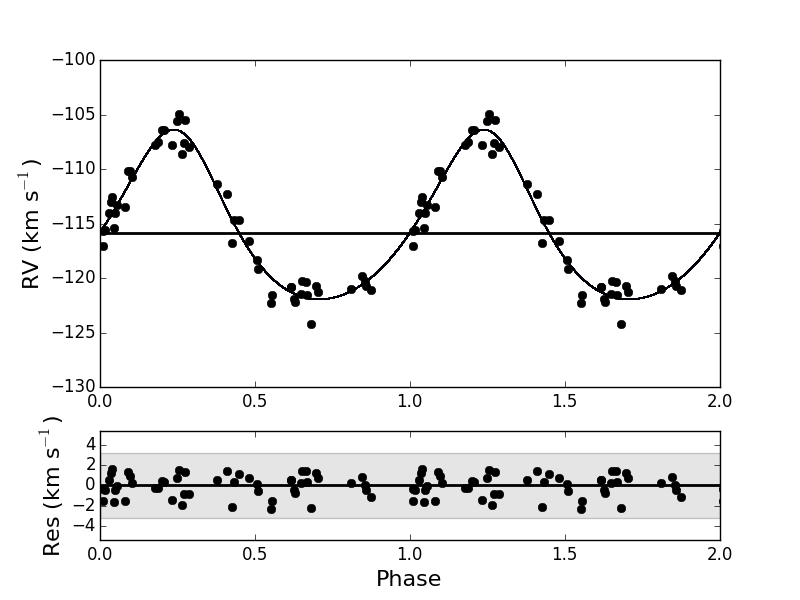}
   \caption{\textit{Top:} Original radial velocity data of HP\,Lyr fitted with a keplerian orbit folded
   on an orbital period of 1650 d and eccentricity of 0.2. \textit{Middle:} Pulsations in the radial velocity folded on a period of
   136 d. The red square symbols and the purple shaded region displays the position in the phase plot where the shocks occur.
    \textit{Bottom:} Pulsation cleaned orbit folded on a period of 1631 d with an eccentricity of 0.17.
     }
     \label{figure:HPLyrorbit}
\end{figure}

\begin{figure}[h]
   \centering
   \includegraphics[width=8cm,height=6cm]{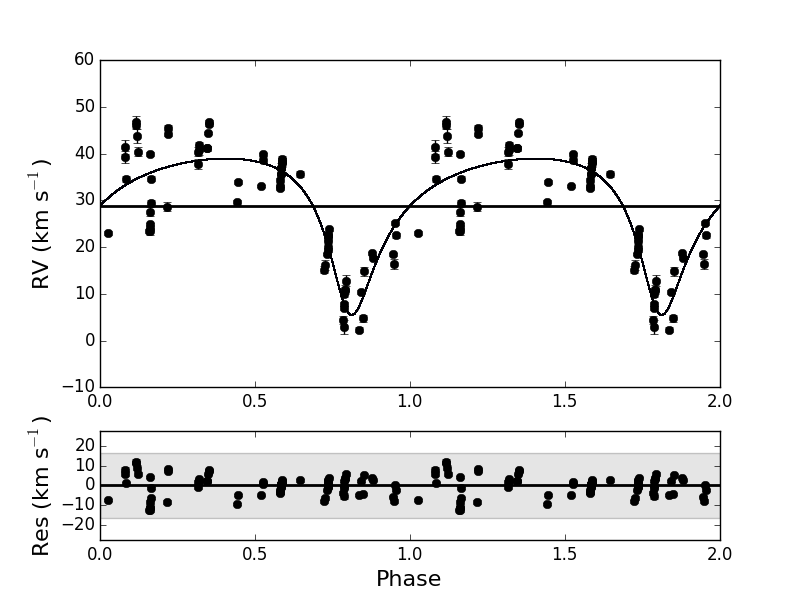}
   \includegraphics[width=8cm,height=6cm]{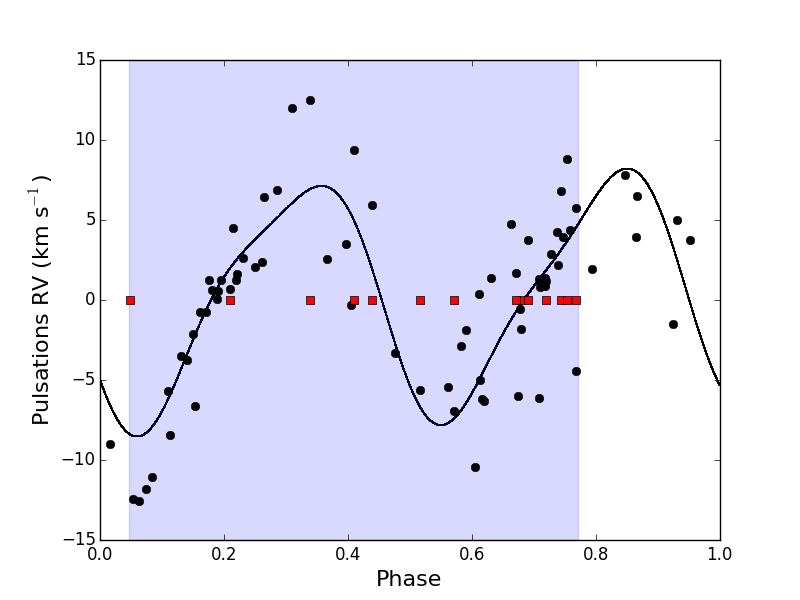}
   \includegraphics[width=8cm,height=6cm]{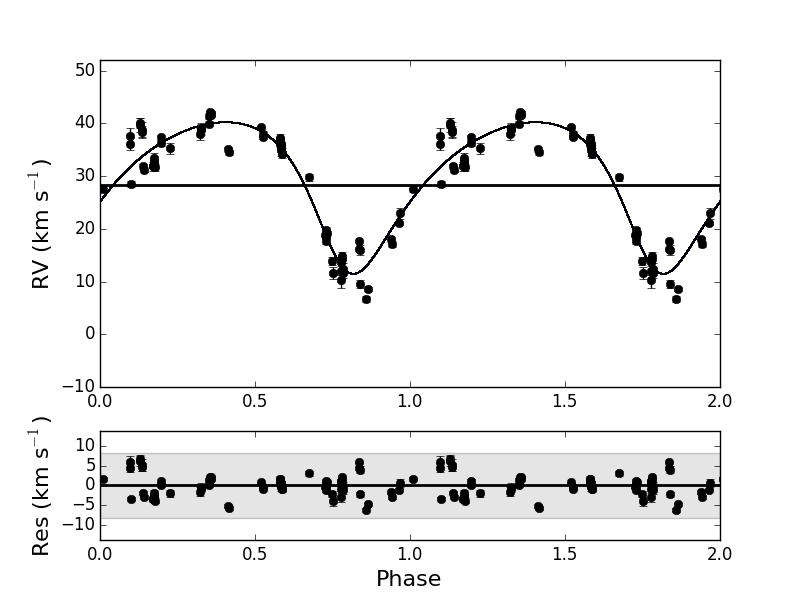}
   \caption{\textit{Top:} Original radial velocity data of IRAS\,09144$-$4933 fitted with a keplerian orbit folded
   on an orbital period of 1640 d and eccentricity of 0.4. \textit{Middle:} Pulsations in the radial velocity folded on a period of
   100 d. The red square symbols and the purple shaded region displays the position in the phase plot where the shocks occur.
    \textit{Bottom:} Pulsation cleaned orbit folded on a period of 1683 d with an eccentricity of 0.45.
     }
     \label{figure:IRAS09144orbit}
\end{figure}

\begin{figure}[h]
   \centering
   \includegraphics[width=8cm,height=6cm]{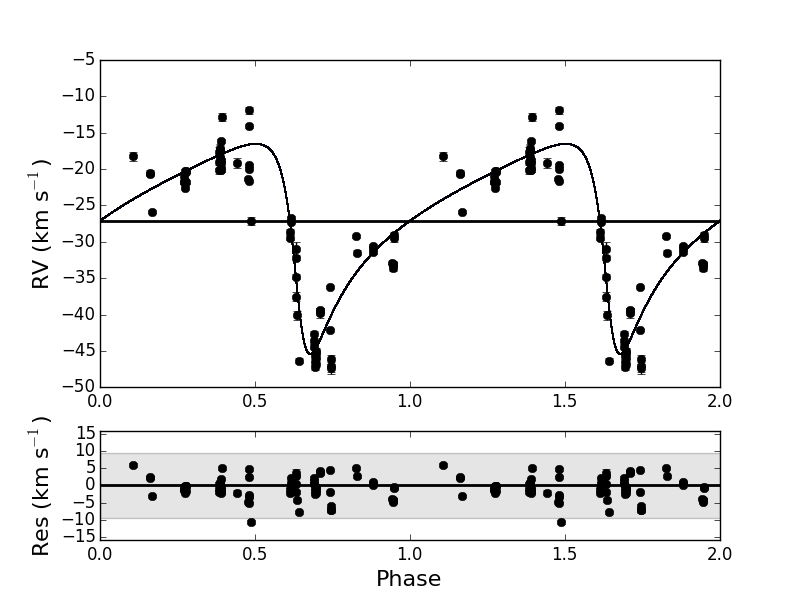}
   \includegraphics[width=8cm,height=6cm]{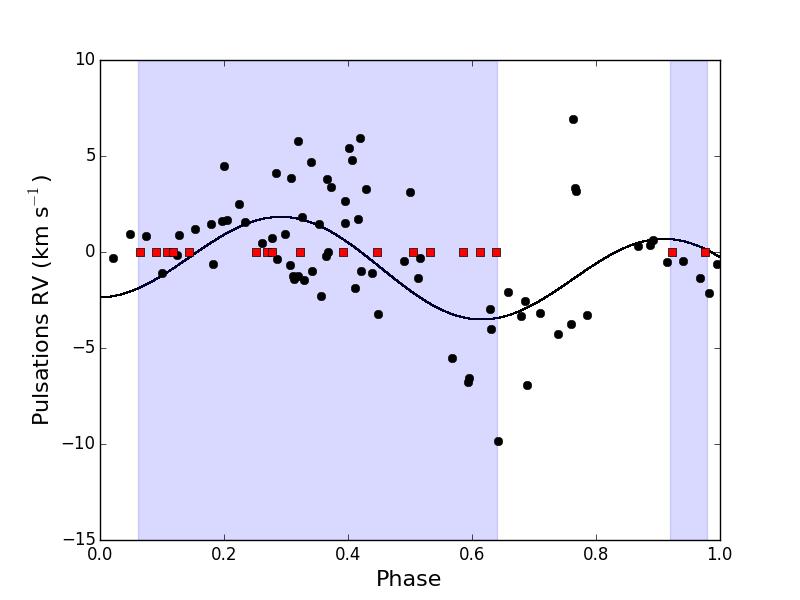}
   \includegraphics[width=8cm,height=6cm]{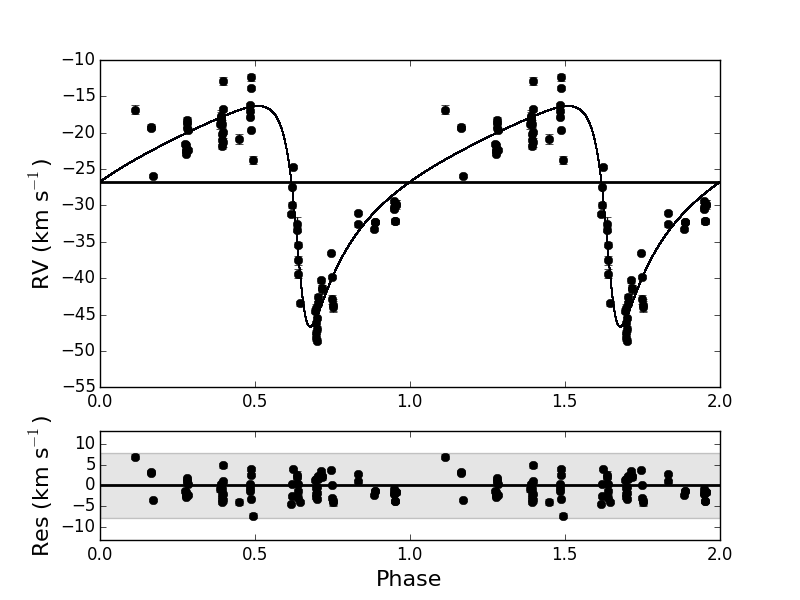}
   \caption{\textit{Top:} Original radial velocity data of IRAS\,17038$-$4815 fitted with a keplerian orbit folded
   on an orbital period of 1393 d and eccentricity of 0.6. \textit{Middle:} Pulsations in the radial velocity folded on a period of
   37.3 d. The red square symbols and the purple shaded region displays the position in the phase plot where the shocks occur.
    \textit{Bottom:} Pulsation cleaned orbit folded on a period of 1386 d with an eccentricity of 0.57.
     }
     \label{figure:IRAS17038orbit}
\end{figure}

The main result is that the six stars in our sample are indeed binaries with orbital periods ranging from 654 to 1683 days. 
We have demonstrated that in the case of strong pulsators, it is a prime factor to clean the radial velocities from the 
pulsation signals in order pin down the orbital parameters as accurately as possible.
The original orbits, the spectroscopic pulsation fit and the pulsation-cleaned orbits are shown in Figs. \ref{figure:DYOriorbit} to \ref{figure:IRAS17038orbit}.
Note that the horizontal black lines in the fit (top and bottom panels of Figs. \ref{figure:DYOriorbit} to \ref{figure:IRAS17038orbit}) are the systemic velocities which 
are outlined in Table \ref{table:orbitalparameters} as $\gamma$. The grey shaded regions in the residuals represent a standard deviation of 3-$\sigma$.
We display the individual binary orbital parameters in Table \ref{table:orbitalparameters}.

\renewcommand{\tabcolsep}{0.2cm}
\begin{table*}[ht]
     \begin{small}
\vspace{0.5ex}
\begin{center}
\begin{tabular}{l|llllllllllll} \hline\hline\rule[0mm]{0mm}{4mm}
			    & DY\,Ori 	&  $\sigma$ & EP\,Lyr& $\sigma$	& HP\,Lyr & $\sigma$	&IRAS\,17038 & $\sigma$ &IRAS\,09144 & $\sigma$ &  TW\,Cam & $\sigma$ \\
  \hline                                                     
P (days) 		    &  1230	& 17  & 1066		&  25	&1631	& 22	&   1386   	& 12  	&  1683	&14	&  654	& 4   	\\
K (km s$^{-1}$) 	    &  	11.7	& 0.4  &  19.0		&  6.4	&7.7	&0.2	&   14.3	& 1.2	&  16.5	&0.5	&  15.7 & 0.5	\\
e                           &  0.21	& 0.04  &  0.6		&  0.06	&0.17	& 0.02	&   0.57	& 0.07 	&  0.45	&0.05	& 0.2  & 0.03	\\
$\omega$ ($^{\circ}$)       & 	1.4	& 0.2  &  1.6		&  0.3	&5.5	& 0.4	&   2.2		& 0.1 	&  2.87	&0.07	&  3.9 & 0.3	\\
T (MJD)       	    	    & 55975 	& 40  & 55158		&  56	&55946	& 48	&   51704	& 17 	& 51416	&24	& 55262 & 29	\\
$\gamma$ (km s$^{-1}$)      &  	-0.1	&  0.3 &  13.4		&  1.6	&-115.1	& 0.1	&   -26.9	&  0.5	&  30.2	&0.3	&  -49.4 &  0.4	\\
$a_1 \sin i$ (AU)           &  1.30	&  0.05 &  1.4		&  0.3	&1.15	& 0.03	&   1.5		&  0.07	&  2.27	&0.05	& 0.93  & 0.03	\\
f(m) (M$_{\odot}$)          &  	0.19	& 0.02  & 0.42 		&  0.04	&0.08	& 0.005	&   0.22	& 0.03 	&  0.55	&0.04	& 0.25  & 0.02	\\
N. Epochs                   &  	71	&   &  58		&  	&65	&	&   75		&  	&  74	&	&  87 	&	\\
\hline
\end{tabular}
\caption{The derived orbital parameters for the pulsation-cleaned orbits. The errors were computed using a Monte-Carlo method on the data points (see text).}
 \label{table:orbitalparameters}
\end{center}  
\end{small}
\end{table*}

\section{Discussion} \label{section:discussion}

\subsection{Luminosities and evolutionary stage} \label{discussion:PLC_dist_lum}
The pulsations of the stars have profitably been used to obtain their luminosities via the PLC relation of the population II cepheids in the LMC.
The big sample of LMC population II cepheids used to derive the PLC relation, allowed us to 
better constrain the uncertainties in the slope and the intercept of the regression, than what is presented in the previous study of \citet{alcock98}. 
We derive the V-band absolute magnitudes as: M$_V$ = $-$2.53\,Log\,(P)\,($\pm$\,0.03) $-$ 1.17\,($\pm$\,0.03) + 2.55\,$\mean{(V - I)_0}$. 
The slope of our PLC; $-$2.53 $\pm$ 0.03, is found to be slightly less steep than the value of $-2.95$ $\pm$ 0.12 given by \citet{alcock98}. 
We also find a slightly more negative value of $-$1.17 $\pm$ 0.03 for the intercept than a value of $-$0.61 $\pm$ 0.2 calculated by \citet{alcock98}.
Our sample of Galactic objects are found in the $\sim$ 2 to 7 kpc vicinity. Using these distances we obtained their luminosities.
Assuming the tip of the RGB to be at a luminosity of $\sim$  2500 L$_\odot$ \citep{kamath2016}, all stars but DY\,Ori and EP\,Lyr, 
can be classified as post-AGB candidates.

Recent surveys in the SMC and LMC enabled the discovery of a new class of dusty objects termed as "post-RGB" objects \citep{kamath2015,kamath2016}. 
These are evolved, dusty objects with mid-IR excesses and stellar parameters mimicking that of post-AGB stars, except that they have lower luminosities 
(100 - 2500 L$_\odot$) than what is expected for a post-AGB star. There are so far 42 of these objects known in the SMC and 119 in the LMC. 
These stars evolve off the RGB much earlier than what we expect by a standard single star evolution scenario. This is very likely the consequence of a strong binary 
interaction process already occurring on the RGB. The stellar parameters and luminosities of DY\,Ori and EP\,Lyr classify them as post-RGB stars. We therefore note that these two stars are the first Galactic counterparts of the newly discovered post-RGB stars in the Magellanic clouds.

\begin{figure}[h!]
   \centering
   \includegraphics[width=9cm,height=7cm]{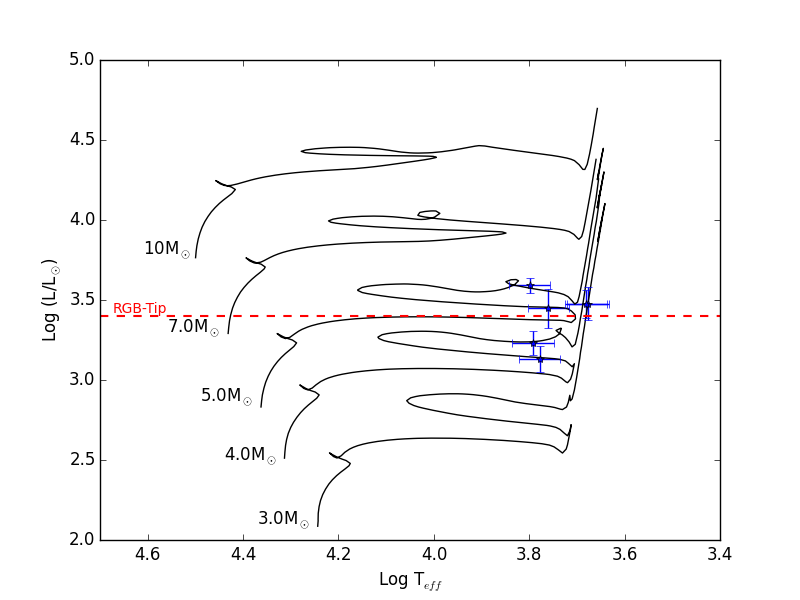} 
   \caption{The position of the Galactic RV Tauri stars on the HRD. The dark lines are evolutionary tracks from \citet{bertelli09}. The red dashed line shows the RGB-tip at 2500\,L$_\odot$. 
   Note that the positions of TW Cam and IRAS\,17038-4815 nearly coincide on this diagram.}
     \label{figure:evo}
\end{figure}

In our sample, DY\,Ori has the lowest luminosity with a pulsation period of 30.4 days. If we consider the RV Tauri stars in the LMC, the lowest pulsation 
period is found to be that of OGLE-LMC-T2CEP-016, with a pulsation period of $\sim$ 20.3 $\pm$ 1 days \citep{soszynsky2008a}. This period translates into a 
luminosity of 1040 $\pm$ 40 L$_\odot$ based on our PLC relation. This is an indication that RV Tauri stars with circumstellar dust can occupy even the low luminosity regime in the HR diagram
making them post-RGB objects (see Fig. \ref{figure:evo}).

 %/home/rajeev/hermesRun/evolution/evo.py

\subsection{The long orbital periods and eccentricites}
\subsubsection{The long orbital periods}
The orbital periods for our sample ranges from 650 days to 1680 days. In Table \ref{table:rlfillingfactors} (Columns 2 and 3, respectively), 
we list for 2 different inclinations of 90$^{\circ}$ and 60$^{\circ}$, the total orbital separation (a$_1$ + a$_2$) at periastron (a$_p$) 
and apastron passage (a$_{ap}$). We used the Eggleton's formula \citep{eggleton83} to estimate the effective roche-lobe radius of the 
primary star at periastron ($R_{RL_p}$) and apastron ($R_{RL_{ap}}$) passage (see Columns 6 and 7). In Column 8 we list the radius of a star 
of the same luminosity, "evolved back" to the RGB or AGB phase, assuming an effective temperature of 3500 K.

The high values of the roche-lobe filling factor shown in the last Column of Table \ref{table:rlfillingfactors} 
suggest that these stars should have interacted due to a RLOF event at an earlier stage of their evolution. More evidence for a RLOF event is 
motivated by circumstellar material observed at this stage of their evolution.  A non-conservative mass transfer event on the AGB, 
causing mass to be transferred through the 2$^{nd}$ and 3$^{rd}$ Lagrangian points is expected in our sample. 
This serves as a logical explanation for the presence of a CB disc. The mentioned evidences suggest that a RLOF scenario 
should have been a likely occurrence in these systems and consequently leading to an orbital shrinkage.

\textit{None of the systems have undergone spiral-in}. It is striking that, despite the evidence for a history of Roche lobe overfilling,
none of the systems encountered a dramatic spiral-in. Such stable RLOF was already proposed by \citet{tout88} and \citet{han95b}. This occurs when the
mass-ratio is reversed prior to RLOF due to a tidally-enhanced wind mass-loss of the primary.

The systems described here are surrounded by likely Keplerian dusty circumstellar discs, which 
store angular momentum \citep[e.g,][]{bujarrabal15}. There is enough observational evidence that inner rim of the disc starts already at the dust
sublimation radius ($\sim$ 3 - 5 AU) for our targets \citep[][]{deruyter06,hillen2016}. The orbital separations shown
in Table 6, suggest that the stellar components are close to the disc inner-rim at apastron passage. This indicates a disc-binary interaction
is very probable, whereby angular momentum can be imparted from the binary system to the disc. It is outside the scope of this paper to investigate this,
but it seems that the presence of a circumbinary disc has an important impact on the orbital evolution of these systems. This will be included in our future research.

\subsubsection{Eccentricities} \label{section:high_ecc}
The eccentricities of the orbits range between 0.2 and 0.6. The non-zero eccentricities are striking. 
Indeed, tides are expected to circularise the orbits very efficiently when the size of the star reaches the Roche Lobe radius \citep[e.g,][]{zahn77,zahn89}. 
The orbital parameters we derive, suggests that the eccentricity is maintained or increased (pumped) by some mechanisms.

A few eccentricity-pumping mechanisms have been proposed to explain the $e-log(P)$ distribution displayed by many long-period binaries with high 
eccentricites. The systems include sdB binaries \citep{vos15}, barium stars \citep{pols03,izzard10}, CH \citep{mcclure90} and CEMP stars \citep{abate15}.
The mechanisms involved are, a tidally-enhanced wind mass-loss \citep{tout88}, an enhanced mass loss at periastron by a RLOF \citep{soker00b},
eccentricity pumping by the CB disc \citep{dermine13} and pumping induced by a wind-RLOF \citep{bonacic2008}. 

An eccentricity-pumping event by the CB disc could possibly explain some of the observed eccentricities in our sample of RV Tauri stars. 
Evidence for a disc-binary interaction is provided by the strongly depleted atmospheres among the RV Tauri stars with a disc \citep[e.g,][]{gezer2015,maas05}. 
The discrepancies between the aforementioned proposed models and the observed eccentricities urges more in-depth studies of 
eccentricity pumping via mass-transfer between the stars, disc-binary interaction and other input physics not yet taken into account.

\subsection{RVb phenomenon, inclinations and nature of companion} \label{discussion:pulsations}

We have used an updated ASAS/AAVSO photometry to study the pulsational variability with an extended 
coverage for the RV Tauri stars. The immediate results include an improved knowledge about their pulsational 
characteristics and brightness modulations throughout the orbit.

We reveal the RVb photometric nature of IRAS\,17038$-$4815, IRAS\,09144$-$4933 and TW\,Cam 
and the RVa photometric type of DY\,Ori. The long-term periodic change in the RVbs (see, Table \ref{table:Oparameters_puls}), of the order of the orbital period provides good evidence 
that the variability is associated to variable extinction through the circumstellar dust. Our results therefore advocates further the arguments of \citet{waelkens91,waelkens93}.
The RVb ``phenomenon'' offers a way to place constraints on the orbital inclinations of the binaries. However, two important factors which needs to be taken into account when predicting the inclinations solely based on their RVb nature are: the inner radius of the disc and the scale-height of the disc.

Generally, the inner radius of post-AGB discs is located at the dust sublimation radius \citep[][Hillen et al. in prep.]{deruyter06,hillen2016}, 
with AC\,Her being the exception \citep[][Hillen et al. in prep.]{hillen2015}. Furthermore, the SEDs indicate that a large amount of the stellar luminosity is 
converted into IR-emission indicating a significant scale-height. In these cases, it could well be possible that the observers' line-of-sight is even grazing 
the disc at lower inclinations than 70$^\circ$. A detailed investigation of these two factors is beyond the scope of this study and we assumed a minimum orbital inclination of 70$^{\circ}$ for the RVb types.

Given that these are wide binaries, the inclinations must be significantly higher than 0$^{\circ}$ to induce a noticeable radial velocity shift.
Also, for inclinations below 35$^\circ$, we find improbably high values for the mass of the companion ($m_2$), under the assumption that the primary is of 0.6\,M$_\odot$ \citep[e.g,][]{weidemann90}. These allow us to put a lower-limit constraint of $\sim$ 35$^\circ$ and an upper limit of 60$^\circ$ for the orbital inclination of the RVas.

\subsubsection{Nature of companion}
The spectra are dominated by the highly luminous primary star and no signature of the companion is seen in the spectra. 
These are therefore single-lined spectroscopic binaries. 

The mass functions range from 0.08 to 0.55 which translates into a minimum mass range of 0.5 to 1.2\,M$_\odot$ (displayed in Table \ref{table:rlfillingfactors}) for the companion, 
assuming a mass of 0.6 M$_\odot$ for the primary and an edge-on inclination. The minimum masses of the companion offers two possibilities of their nature; 
either a low-luminosity white dwarf (WD) or an unevolved low-mass main sequence star.

We argue that the companion is likely not a WD for two reasons:
\begin{itemize}
 \item Considering the mass of a white dwarf is well constrained 
at $\sim$ 0.6 $\pm$ 0.1\,M$_\odot$ \citep[e.g,][]{weidemann90}. Our lower mass limit of the companion almost rules out this possibility.
Note here, HP\,Lyr still qualifies to have a WD secondary based on a minimum companion mass of $\sim$ 0.5\,M$_\odot$. 
\item A compact white dwarf companion would have accreted mass from the primary causing the system to most probably show symbiotic activity. 
\end{itemize}

\begin{table*}
\begin{center}
\begin{small}
\begin{tabular}{l|ccccccccc}
 \hline\hline                
			&	& &	&& \centering {\bfseries $\bf i$ = 90$^{\bf \circ}$} &	 	&   & \\ \hline
   \centering Star	& \centering $a_p$ (AU)	& \centering $a_{ap}$ (AU) &  \centering $m_2$ (M$_{\odot}$) & \centering $R$ (R$_{\odot}$)  & \centering $R_{RL_p}$ (R$_{\odot}$) 	& \centering $R_{RL_{ap}}$ (R$_{\odot}$) 	& \centering $R_{RGB/AGB}$ (R$_{\odot}$)	& \centering $f$  & \\ \hline   
 DY\,Ori 	 &2.0 &3.0 &0.7 & 40& 156& 238&100& 0.4-0.6& \\ 
 EP\,Lyr 	 &0.9 &3.8 &0.7 & 36& 74 & 297&112& 0.4-1.5& \\ 
 HP\,Lyr 	 &2.5 &3.5 &0.5 & 60& 214& 302&170& 0.6-0.8& \\ 
 IRAS\,17038 	 &1.2 &4.3 &0.8 & 73& 91 & 331&149& 0.5-1.6& \\ 
 IRAS\,09144 	 &1.9 &4.9 &1.2 & 58& 129& 340&144& 0.4-1.1& \\ 
 TW\,Cam 	 &1.3 &1.9 &0.7 & 83& 102 & 153&149& 1.0-1.5& \\ \hline \hline
   			&	&	&	&&\centering \bfseries $\bf i$ = 60$^{\bf \circ}$	&  			&	&  & \\ \hline
 DY\,Ori 	 & 1.8 &2.7 &0.9 & 40& 132 & 203& 100& 0.5-0.8& \\ 
 EP\,Lyr 	 & 0.8 &3.4 &0.9 & 36& 63  & 253& 112& 0.4-1.8& \\ 
 HP\,Lyr 	 & 2.2 &3.1 &0.6 & 60& 181 & 255& 170& 0.7-0.9& \\ 
 IRAS\,17038 	 & 1.1 &3.9 &1.0 & 73& 77 & 281& 149& 0.5-1.9& \\ 
 IRAS\,09144 	 & 1.7 &4.5 &1.6 & 58& 109 & 289& 144& 0.5-1.3& \\ 
 TW\,Cam 	 & 1.2 &1.7 &0.8 & 83& 87 & 131& 149& 1.1-1.7& \\ \hline

\end{tabular}
\caption{The orbital separation at periastron and apastron passage is shown in Columns 2 and 3, respectively for two different inclinations of 90$^{\circ}$ and 60$^{\circ}$. In Column 4 we show 
the estimated masses of the companion assuming a primary mass of 0.6 M$_{\odot}$.
Column 5 is the current radius of the primary derived from the effective temperature and luminosity. The 6$^{th}$ and 7$^{th}$ Columns display the radius of the Roche-Lobe at periastron and apastron, respectively
. The 8$^{th}$ Column outlines the estimated radius of the star on the RGB/AGB, assuming an effective temperature of T$_{RGB/AGB}$ $\sim$ 3500 K. We show the Roche-Lobe filling factors at apastron-periastron 
($f$ = $R_{AGB}$/$R_{RL}$ or $f$ = $R_{RGB}$/$R_{RL}$) in the last Column.}
 \label{table:rlfillingfactors}
\end{small}
\end{center}
\end{table*}

\section{Conclusions}   \label{section:conclusion}  

We have used high-resolution spectroscopic data from the HERMES and CORALIE spectrographs to probe the binary status of a sample of 
6 Galactic RV Tauri stars with a disc. All of them are binaries and we therefore extend the sample of known binary RV Tauri stars with a disc from 6 to now 12. 

We present a novel PLC relationship calibrated using data for 187 type II cepheids in the LMC, available from \citet{soszynsky2008a}. 
The PLC was used to obtain accurate distances and hence luminosities of our sample of Galactic RV Tauri stars. The luminosities of 
HP\,Lyr, IRAS\,17038$-$4815, IRAS\,09144$-$4933 and TW Cam, suggests that they are post-AGB stars. The relatively lower luminosities 
of DY\,Ori and EP\,Lyr shows that these objects have evolved off the RGB, making them dusty post-RGB objects. According to theory, 
these stars should have experienced a spiral-in event and have a circularised orbit at this stage. The observed long orbital periods 
and eccentricities indicates that these binary systems do not follow standard binary evolution models.

The stars included in this study, show mild-to-highly depleted photospheres; a consequence of re-accretion of gas from the disc.
Together with the disc-binarity correlation, the link between binarity and depletion is extended.

Long-term variations in the photometric time series of the order of the orbital period is detected for IRAS\,17038$-$4815, IRAS\,09144$-$4933 and TW\,Cam. 
This enabled us to classify them as RVb photometric types and consequently place constraints on their orbital inclinations and the mass of their companions. 
We also prove the RVa nature of DY\,Ori due to no significant long-term variation seen in its photometric time series.

\begin{acknowledgements} 
This research has been carried out based on observations made using the Mercator Telescope, operated on the island of La Palma by the Flemish Community, 
situated at the Spanish Observatorio del Roque de los Muchachos of the Instituto de Astro\'fisica de Canarias. We used data from the HERMES spectrograph, 
supported by the Fund for Scientific Research of Flanders (FWO), the Research Council of K.U. Leuven, the Fonds National 
Recherches Scientific (FNRS), the Royal Observatory of Belgium, the Observatoire de Gen\`{e}ve, Switzerland and the Thringer Landessternwarte Tautenburg, Germany. 
This research has been conducted based on funding from the Research Council of K.U. Leuven and was partially funded by the Belgian Science Policy Office under contract BR/143/A2/STARLAB.
In this study we also included data from the Swiss Leonhard Euler
1.2\,m Telescope located at the Haute Provence Observatory in La Silla. We used the following internet-based resources: NASA Astrophysics Data System for bibliographic services, Simbad,
VizieR online catalogue operated by CDS, ASAS and AAVSO photometric databases. RM thanks Brent Miszalski, Shazrene Mohamed, Peter Wood and Jonas Debosscher for their helpful comments and inputs.
Lastly, the authors thank the anonymous referee for the valuable comments which helped in improving the paper.
           
\end{acknowledgements}

\bibliographystyle{aa}          
\bibliography{allreferences}

\appendix
\section{Spectral Energy Distributions} \label{appendix:AppendixA}

\newpage

\begin{figure}
   \centering
   \includegraphics[width=8cm,height=6cm]{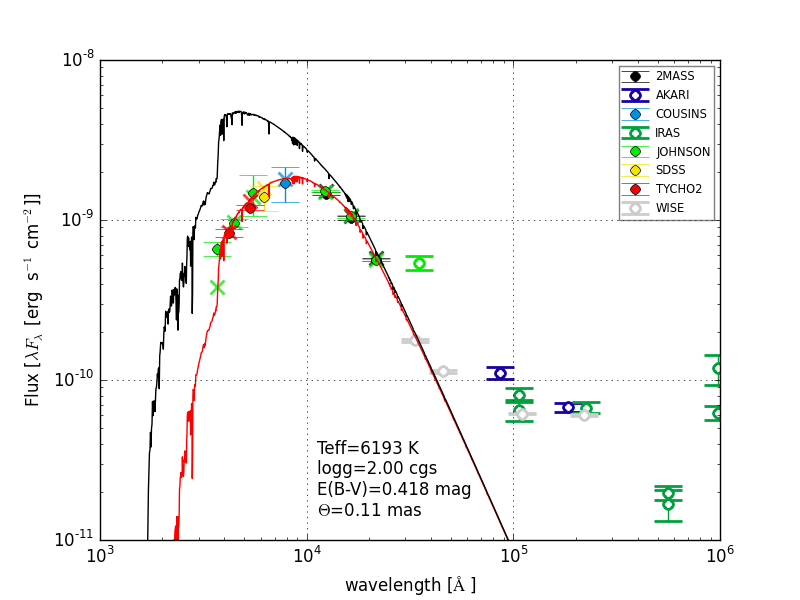} 
   \caption{The SED of EP\,Lyr. The red model is the stellar photosphere fitted using a Kurucz model and the black 
   line is the dereddened model.}
     \label{figure:SEDEP_Lyr}
\end{figure} %/home/rajeev/hermesRun/pulsations/ASASPhotometry/allplots_time.py

\begin{figure}
   \centering
   \includegraphics[width=8cm,height=6cm]{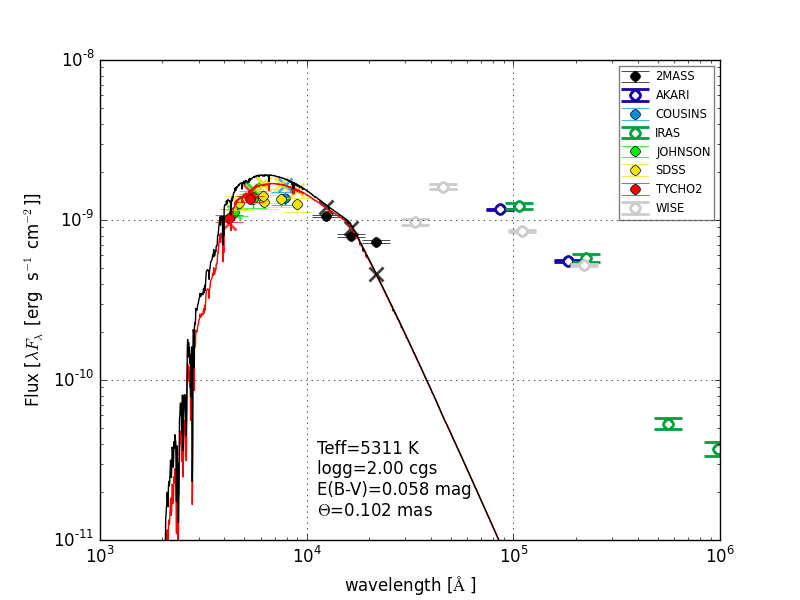} 
   \caption{The SED of HP\,Lyr. The red model is the stellar photosphere fitted using a Kurucz model and the black 
   line is the dereddened model.}
     \label{figure:SEDHPLyr}
\end{figure} %/home/rajeev/hermesRun/pulsations/ASASPhotometry/allplots_time.py

\begin{figure}
   \centering
   \includegraphics[width=8cm,height=6cm]{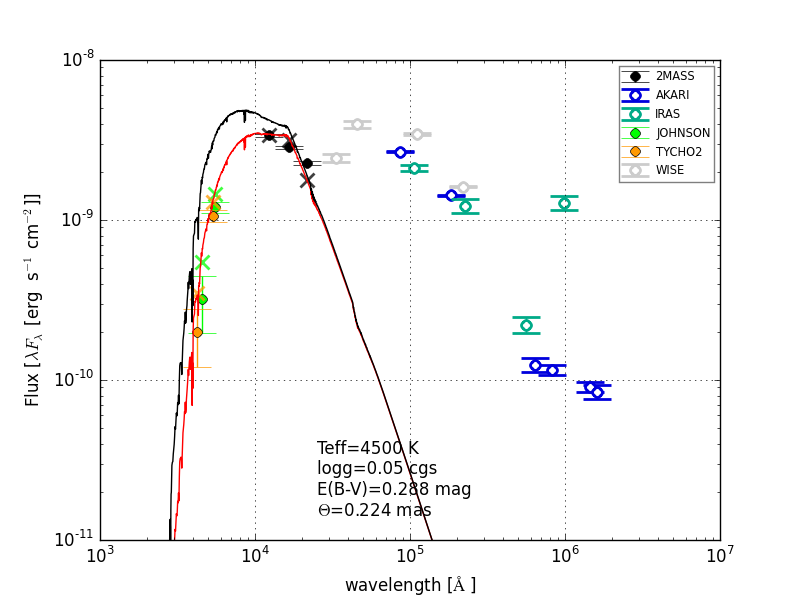} 
   \caption{The SED of IRAS\,17038$-$4815. The red model is the stellar photosphere fitted using a Kurucz model and the black 
   line is the dereddened model.}
     \label{figure:SEDIRAS17038}
\end{figure} %/home/rajeev/hermesRun/pulsations/ASASPhotometry/allplots_time.py

\begin{figure}
   \centering
   \includegraphics[width=8cm,height=6cm]{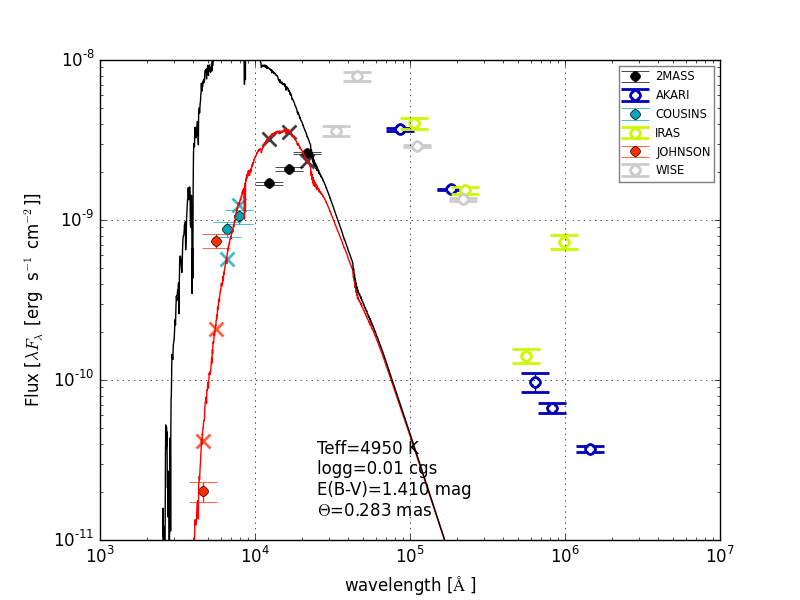} 
   \caption{The SED of IRAS\,09144$-$4933. The red model is the stellar photosphere fitted using a Kurucz model and the black 
   line is the dereddened model.}
     \label{figure:SEDIRAS09144}
\end{figure} %/home/rajeev/hermesRun/pulsations/ASASPhotometry/allplots_time.py

\begin{figure}
   \centering
   \includegraphics[width=8cm,height=6cm]{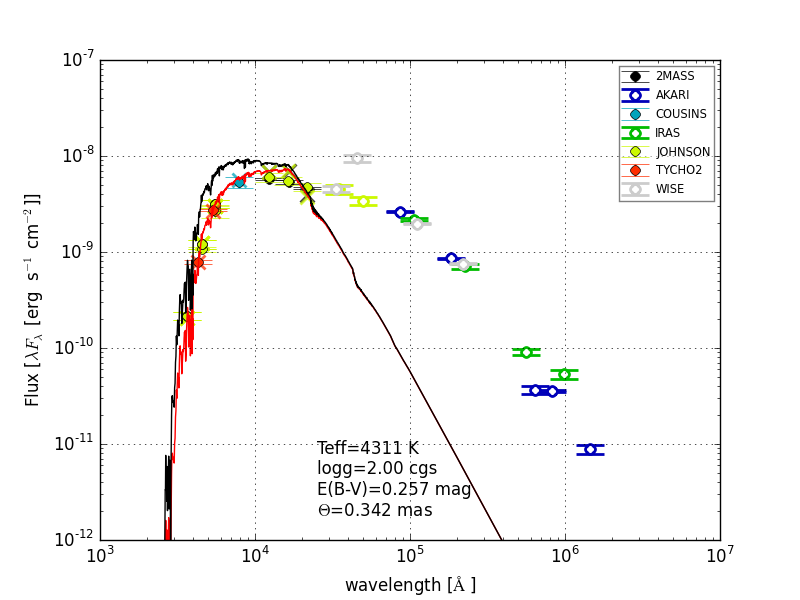} 
   \caption{The SED of TW\,Cam. The red model is the stellar photosphere fitted using a Kurucz model and the black 
   line is the dereddened model.}
     \label{figure:SEDTWCam}
\end{figure} %/home/rajeev/hermesRun/pulsations/ASASPhotometry/allplots_time.py

\newpage
\section{SED Photometry} \label{appendix:AppendixB}

\begin{table}
\tiny                             
\begin{center}
\begin{tabular}{llll}   \\ \hline \hline
Phot. Band & Wavelength ($nm$) & Magnitude (mag) & Error (mag) \\ \hline
2MASS.J &   1239.31 & 8.073 & 0.023 \\
2MASS.H &   1649.49 & 7.381 & 0.017 \\
2MASS.KS &  2163.86 & 6.983 & 0.018 \\
IRAS.F12 &  11035.7 & 12.4 & 1.36 \\
IRAS.F25 &  23072.4 & 14.9 & 1.49 \\
IRAS.F60 &  58190.4 & 4.18 & 0.37 \\
IRAS.F12 &  11035.7 & 12.87 & 0.64 \\
IRAS.F25 &  23072.4 & 14.87 & 1.13 \\
IRAS.F60 &  58190.4 & 3.974 & 0.270 \\
IRAS.F100 & 99519.3 & 21.85 & 5.26 \\
JOHNSON.V & 549.292 & 11.639 & 0.243 \\
COUSINS.I & 787.479 & 9.766 & 0.216 \\
JOHNSON.V & 549.292 & 11.59 & - \\
JOHNSON.B & 443.354 & 13.11 & - \\
JOHNSON.U & 363.73  & 14.23 & - \\
SDSS.RP &   624.72  & 10.841 & 0.109 \\
AKARI.S9W & 8852.03 & 8.363 & 0.123 \\
AKARI.L18W& 18918.6 & 12.98 & 0.23 \\ \hline
\end{tabular}
\caption{Table showing the different photometric bands used to construct the SED of DY\,Ori.}  
\label{table:dyori_sedphot}
\end{center}
\end{table}

\begin{table}
\tiny                           
\begin{center}
\begin{tabular}{llll}  \label{table:eplyr_sedphot} \\ \hline \hline
Phot. Band & Wavelength ($nm$) & Magnitude (mag) & Error (mag) \\ \hline
2MASS.J &   1239.31 & 8.534 & 0.023 \\
2MASS.H &   1649.49 & 8.153 & 0.027 \\
2MASS.KS &  2163.86 & 8.03 & 0.024 \\
IRAS.F12 &  11035.7 & 0.306 & - \\
IRAS.F60 &  58190.4 & 0.4 & - \\
IRAS.F100 & 99519.3 & 2.1 & - \\
IRAS.F12 &  11035.7 & 0.2482 & 0.0374 \\
IRAS.F25 &  23072.4 & 0.5301 & 0.0408 \\
IRAS.F60 &  58190.4 & 0.3399 & 0.0744 \\
IRAS.F100 & 99519.3 & 3.982 & 0.856 \\
TYCHO2.BT & 419.631 & 11.374 & 0.055 \\
TYCHO2.VT & 530.699 & 10.635 & 0.042 \\
JOHNSON.B & 443.354 & 11.168 & 0.057 \\
JOHNSON.V & 549.292 & 10.334 & 0.314 \\
COUSINS.I & 787.479 & 9.365 & 0.265 \\
JOHNSON.L & 3527.07 & 6.6 & - \\
JOHNSON.U & 363.73 & 10.92 & - \\
JOHNSON.J & 1243.06 & 8.53 & 0.01 \\
JOHNSON.H & 1644.17 & 8.15 & 0.02 \\
JOHNSON.K & 2187.92 & 8.03 & 0.02 \\
SDSS.RP &   624.72 & 10.244 & 0.195 \\
AKARI.S9W & 8852.03 & 0.3357 & 0.0287 \\
AKARI.L18W &18918.6 & 0.438 & 0.0301 \\
WISE.W1 &   3345.97 & 7.969 & 0.024 \\
WISE.W2 &   4595.22 & 7.474 & 0.02 \\
WISE.W3 &   11548.5 & 5.167 & 0.014 \\
WISE.W4 &   22078.9 & 3.172 & 0.019 \\ \hline
\end{tabular}
\caption{Table showing the different photometric bands used to construct the SED of EP\,Lyr.}    
\end{center}
\end{table}

\begin{table}
\tiny                           
\begin{center}
\begin{tabular}{llll}  \label{table:hplyr_sedphot} \\ \hline \hline
Phot. Band & Wavelength ($nm$) & Magnitude (mag) & Error (mag) \\ \hline
2MASS.J &   1239.31 & 8.877 & 0.02 \\
2MASS.H &   1649.49 & 8.429 & 0.02 \\
2MASS.KS &  2163.86 & 7.75 & 0.02 \\
IRAS.F12 &  11035.7 & 4.66 & 0.18 \\
IRAS.F25 &  23072.4 & 4.56 & 0.27 \\
IRAS.F60 &  58190.4 & 1.08 & 0.09 \\
IRAS.F100 & 99519.3 & 1.25 & - \\
TYCHO2.BT & 419.631 & 11.147 & 0.051 \\
TYCHO2.VT & 530.699 & 10.497 & 0.042 \\
JOHNSON.B & 443.354 & 11.003 & 0.052 \\
JOHNSON.V & 549.292 & 10.402 & 0.043 \\
JOHNSON.V & 549.292 & 10.43 & 0.13 \\
COUSINS.I & 787.479 & 9.608 & 0.094 \\
SDSS.RP &   624.72  & 10.317 & - \\
AKARI.S9W & 8852.03 & 3.545 & 0.0266 \\
AKARI.L18W& 18918.6 & 3.602 & 0.0467 \\
SDSS.G &    467.745 & 10.652 & - \\
SDSS.R &    616.825 & 10.239 & - \\
SDSS.I &    749.567  & 10.08 & - \\
SDSS.Z &    896.965  & 9.964 & - \\
WISE.W1 &   3345.97  & 6.118 & 0.051 \\
WISE.W2 &   4595.22  & 4.585 & 0.044 \\
WISE.W3 &   11548.5  & 2.31 & 0.013 \\
WISE.W4 &   22078.9  & 0.825 & 0.014 \\ \hline
\end{tabular}
\caption{Table showing the different photometric bands used to construct the SED of HP\,Lyr.}    
\end{center}
\end{table}

\begin{table}
\tiny                           
\begin{center}
\begin{tabular}{llll}  \label{table:ir17038_sedphot} \\ \hline \hline
Phot. Band & Wavelength ($nm$) & Magnitude (mag) & Error (mag) \\ \hline
2MASS.J &     1239.31 & 7.622 & 0.017 \\
2MASS.H &     1649.49 & 7.05 & 0.027 \\
2MASS.KS &    2163.86 & 6.523 & 0.027 \\
IRAS.F12 &    11035.7 & 8.05 & 0.40 \\
IRAS.F25 &    23072.4 & 9.67 & 0.96 \\
IRAS.F60 &    58190.4 & 4.49 & 0.4939 \\
IRAS.F100 &   99519.3 & 43.3 & - \\
TYCHO2.BT &   419.631 & 12.934 & 0.425 \\
TYCHO2.VT &   530.699 & 10.774 & 0.094 \\
JOHNSON.B &   443.354 & 12.371 & 0.425 \\
JOHNSON.V &   549.292   & 10.567 & 0.093 \\
AKARI.N60 &   65398.6 & 2.758 & - \\
AKARI.WIDES & 85124.0 & 3.42 & 0.24 \\
AKARI.WIDEL & 1.46434e+05 & 4.442 & 0.313 \\
AKARI.N160 &  1.61669e+05 & 4.534 & - \\
AKARI.S9W &   8852.03 & 8.118 & 0.0541 \\
AKARI.L18W &  18918.6 & 9.277 & 0.0957 \\
WISE.W1 &     3345.97 & 5.113 & 0.061 \\
WISE.W2 &     4595.22 & 3.612 & 0.055 \\
WISE.W3 &     11548.5 & 0.801 & 0.009 \\
WISE.W4 &     22078.9 & -0.398 & 0.008 \\ \hline
\end{tabular}
\caption{Table showing the different photometric bands used to construct the SED of IRAS\,17038$-$4815.}    
\end{center}
\end{table}

\begin{table}
\tiny                           
\begin{center}
\begin{tabular}{llll}  \label{table:ir09144_sedphot} \\ \hline \hline
Phot. Band & Wavelength ($nm$) & Magnitude (mag) & Error (mag) \\ \hline
2MASS.J &      1239.31 & 8.372 & 0.026 \\
2MASS.H &      1649.49 & 7.391 & 0.036 \\
2MASS.KS &     2163.86 & 6.369 & 0.017 \\
IRAS.F12 &     11035.7 & 15.3 & 1.22 \\
IRAS.F25 &     23072.4 & 12.1 & 0.60 \\
IRAS.F60 &     58190.4 & 2.86 & 0.29 \\
IRAS.F100 &    99519.3 & 24.7 & - \\
AKARI.N60 &    65398.6 & 2.152 & 0.283 \\
AKARI.WIDES &  85124.0 & 1.968 & 0.143 \\
AKARI.WIDEL &  1.46434e+05 & 1.83 & 0.09 \\
AKARI.S9W &    8852.03 & 11.17 & 0.234 \\
AKARI.L18W &   18918.6 & 10.08 & 0.0792 \\
WISE.W1 &      3345.97 & 4.698 & 0.074 \\
WISE.W2 &      4595.22 & 2.855 & 0.07 \\
WISE.W3 &      11548.5 & 0.983 & 0.015 \\
WISE.W4 &      22078.9 & -0.201 & 0.021 \\
JOHNSON.B &    443.354 & 15.886 & 0.154 \\
JOHNSON.V &    549.292 & 13.89 & 0.111 \\
COUSINS.R &    549.292 & 12.293 & 0.122 \\
COUSINS.I &    549.292 & 10.797 & 0.107 \\ \hline
\end{tabular}              
\caption{Table showing the different photometric bands used to construct the SED of IRAS\,09144$-$4933.}    
\end{center}
\end{table}

\begin{table} 
\tiny                           
\begin{center}
\begin{tabular}{llll}  \\ \hline \hline
Phot. Band & Wavelength ($nm$) & Magnitude (mag) & Error (mag) \\ \hline
2MASS.J &   1239.31 & 7.035 & 0.02 \\
2MASS.H &   1649.49 & 6.364 & 0.049 \\
2MASS.KS &  2163.86 & 5.75 & 0.017 \\
IRAS.F12 &  11035.7 & 8.25 & 0.33 \\
IRAS.F25 &  23072.4 & 5.6 & 0.336 \\
IRAS.F60 &  58190.4 & 1.84 & 0.1472 \\
IRAS.F100 & 99519.3 & 1.79 & 1 \\
TYCHO2.BT & 419.631 & 11.445 & 0.061 \\
TYCHO2.VT & 530.699 & 9.731 & 0.023 \\
JOHNSON.B & 443.354 & 11.078 & 0.067 \\
JOHNSON.V & 549.292 & 9.586 & 0.025 \\
JOHNSON.V & 549.292 & 9.7 & 0.174 \\
COUSINS.I & 787.479 & 8.134 & 0.134 \\
JOHNSON.V & 549.292 & 9.51 & - \\
JOHNSON.U & 363.73 & 12.12 & - \\
JOHNSON.B & 443.354 & 10.94 &- \\
JOHNSON.J & 1243.06 & 7.04 & - \\
JOHNSON.H & 1644.17 & 6.36 & - \\
JOHNSON.K & 2187.92 & 5.7 &  - \\
JOHNSON.L & 3527.07 & 4.3 &  - \\
JOHNSON.M & 5011.62 & 3.5 &  - \\
JOHNSON.V & 549.292 & 9.51 & - \\
JOHNSON.B &   443.354 & 10.94& - \\
JOHNSON.U &   363.73 & 12.12 & - \\
AKARI.N60 &   65398.6 & 0.815 & - \\
AKARI.WIDES & 85124.0     & 1.057 & 0.0249 \\
AKARI.WIDEL & 1.46434e+05 & 0.4321 & - \\
AKARI.S9W &   8852.03 & 7.969 & 0.0913 \\
AKARI.L18W &  18918.6 & 5.566 & 0.0319 \\
WISE.W1 &     3345.97 & 4.452 & 0.08 \\
WISE.W2 &     4595.22 & 2.66 & 0.087 \\
WISE.W3 &     11548.5 & 1.403 & 0.007 \\
WISE.W4 &     22078.9 & 0.424 & 0.016 \\
JOHNSON.V &   549.292 & 9.51 & - \\
JOHNSON.B &   443.354 & 10.94 & - \\
JOHNSON.U &   363.73  & 12.12 & - \\ \hline

\end{tabular}
\caption{Table showing the different photometric bands used to construct the SED of TW\,Cam.}  
\label{table:twcam_sedphot} 
\end{center}
\end{table}

\end{document}